\documentclass[sigconf]{acmart}

\AtBeginDocument{%
  \providecommand\BibTeX{{%
    \normalfont B\kern-0.5em{\scshape i\kern-0.25em b}\kern-0.8em\TeX}}}

\usepackage{hyperref}
\usepackage{multirow}
\usepackage{graphicx}
\usepackage{graphics}
\usepackage{subcaption}
\usepackage{amsmath}
\usepackage{algorithm}
\usepackage{algorithmic}
\usepackage{wrapfig}
\usepackage{caption}
\usepackage{float}
\usepackage[normalem]{ulem}
\useunder{\uline}{\ul}{}
\usepackage{bbding}
\usepackage{pifont}
\usepackage{amsthm}
\theoremstyle{definition}
\newtheorem{definition}{Definition}[section]

\begin{document}

\title{Joint Identifiability of Cross-Domain Recommendation via Hierarchical Subspace Disentanglement}


\author{Jing Du}
\authornote{Jing Du and Zesheng Ye share equal contributions as co-first authors.}
\authornote{Work done during Jing Du's studies at the University of New South Wales.}
\affiliation{
    \institution{Macquarie University}
    \city{Sydney}
    \country{Australia}
}
\email{jing.du@mq.edu.au}

\author{Zesheng Ye}
\authornotemark[1]
\affiliation{
    \institution{The University of New South Wales}
    \city{Sydney}
    \country{Australia}
}
\email{zesheng.ye@unsw.edu.au}

\author{Bin Guo}
\affiliation{
    \institution{Northwestern Polytechnical University, China}
    \city{}
    \country{}
}
\email{guobin.keio@gmail.com}

\author{Zhiwen Yu}
\affiliation{
    \institution{Northwestern Polytechnical University, China}
    \city{}
    \country{}
}
\email{zhiweny@gmail.com}

\author{Lina Yao}
\authornote{Lina Yao is also affiliated with the University of New South Wales and Macquarie University.}
\affiliation{
    \institution{CSIRO's Data 61}
    \city{Sydney}
    \country{Australia}
}
\email{lina.yao@unsw.edu.au}
\renewcommand{\shortauthors}{Jing Du, et al.}

\begin{abstract}

Cross-Domain Recommendation~(CDR) seeks to enable effective knowledge transfer across domains. 
Most existing works rely on either representation alignment or transformation bridges, but they come with shortcomings regarding {\it identifiability} of domain-shared and domain-specific latent factors.
Specifically, while CDR describes user representations as a joint distribution over two domains, these methods fail to account for its {\it joint identifiability} as they primarily fixate on the marginal distribution within a particular domain.
Such a failure may overlook the conditionality between two domains and how it contributes to latent factor disentanglement, leading to negative transfer when domains are weakly correlated.
In this study, we explore what {\it should} and {\it should not be transferred} in cross-domain user representations from a causality perspective.
We propose a \textbf{H}ierarchical subspace disentanglement approach to explore the \textbf{J}oint \textbf{ID}entifiability of cross-domain joint distribution, termed \textbf{HJID}, to preserve domain-specific behaviors from domain-shared factors.
HJID abides by the feature hierarchy and divides user representations into generic shallow subspace and domain-oriented deep subspaces.
We first encode the generic pattern in the shallow subspace by minimizing the Maximum Mean Discrepancy of initial layer activation.
Then, to dissect how domain-oriented latent factors are encoded in deeper layers activation, we construct a cross-domain causality-based data generation graph, which identifies cross-domain consistent and domain-specific components, adhering to the Minimal Change principle.
This allows HJID to maintain stability whilst discovering unique factors for different domains, all within a generative framework of invertible transformations that guarantee the {\it joint identifiability}.
With experiments on real-world datasets, we show that HJID outperforms SOTA methods on a range of strongly and weakly correlated CDR tasks.
\end{abstract}




\maketitle

\section{Introduction}
\begin{figure}
    \centering
    \includegraphics[width=\linewidth]{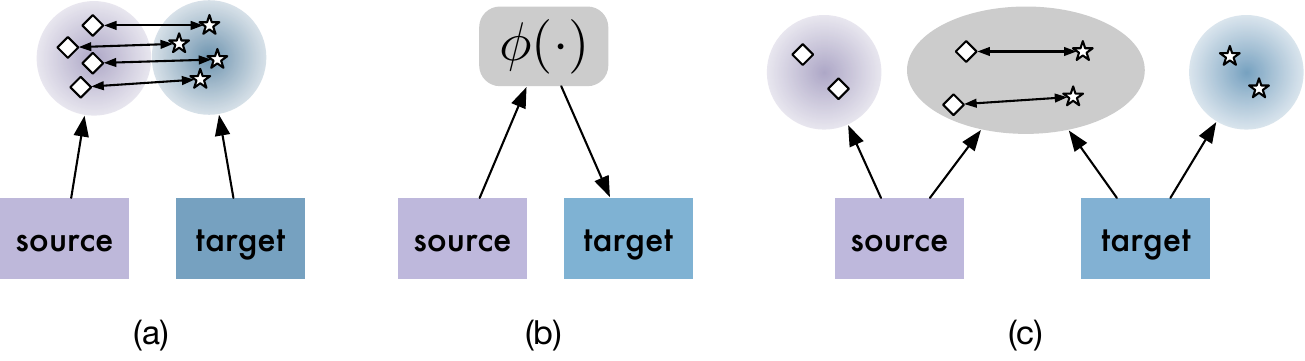}
    \caption{(a) Alignment method; (b) Bridge method; (c) Constraint method.}
    \vspace{-7mm}
    \label{fig:example}
\end{figure}

Cross-Domain Recommendation~(CDR) leverages information from the source domain to the target domain to achieve robust predictions in both domains\cite{cantador2015cross, khan2017cross}.
Let $\boldsymbol{U}_{x}$ and $\boldsymbol{U}_{y}$ denote user representations for the source domain and the target domain, each associated with a domain identifier $\{ X, Y \}$.
A probabilistic viewpoint interprets CDR methods by modeling a joint distribution $P(\boldsymbol{U}_{x}, \boldsymbol{U}_{y})$ with a representation generator capable of optimally inferring user representations $\boldsymbol{U}_{y}$ in the target domain, using observed source user representations $\boldsymbol{U}_{x}$, and vice versa.
From this joint distribution, it becomes feasible to deduce marginal distributions describing user behaviors for each domain, specifically $P(\boldsymbol{U}_{x})$ and $P(\boldsymbol{U}_{y})$.

With this formulation, we can partition previous CDR studies into three pivotal categories, as shown in Fig. \ref{fig:example}: 
\textbf{Alignment method}, \textbf{Bridge method}, and \textbf{Constraint method}.
The \textbf{Alignment} method, featured in~\cite{singh2008relational, wang2021low, yu2022cross}, employs either explicit sharing or implicit alignment of users feature spaces across domains.
These methods enforce $P(\boldsymbol{U}_x ) = P(\boldsymbol{U}_y)$, thus facilitating the consistency of user representations in both domains.
The \textbf{Bridge} method assumes a user's representation can be transferred from one source to another via bespoke neural architectures~\cite{man2017cross, kang2019semi, zhu2021transfer}, which craft a function $\phi(\cdot)$ to map the marginal distribution from one to another $P(\boldsymbol{U}_y ) = \phi(P(\boldsymbol{U}_x))$.
These methods, however, consider user representation as a whole and are fully transferable, potentially leading to negative transfer if improper transformations are applied to non-transferable components~\cite{zhu2021cross}.
To address this, recent studies have veered towards \textbf{Constraints} method~\cite{cao2022disencdr, cao2022cross, du2023distributional}.
These studies emphasize identifying domain-shared components from user representations, and constraining their consistency across domains.

Nevertheless, even {\bf Constraints} methods tend to focus overwhelmingly on extracting domain-shared, transferable features~\cite{cao2022cross, cao2022disencdr, 10.1145/3488560.3498388}, implicitly assuming the remaining, i.e., domain-specific representations are mutually exclusive for different domains.
That is, the domain-specific latent factors for $X$ have no influence on the user behavior in $Y$, and vice versa.
However, from a user preference viewpoint, $\boldsymbol{U}_{x}$ and $\boldsymbol{U}_{y}$ inherently suggest the user's intrinsic preference shifts to different domains $X$ and $Y$, but not necessarily imply they are independent quantities~\cite{stojanov2019data}, otherwise it might be overly restrictive.
To illustrate, we assume $\boldsymbol{U}$ is a Gaussian random variable in the simplest case, whose mean can be viewed as domain-shared factors arise from behavioral consistency, whereas domain-specific factors deliver observation variances owing to distribution shifts across domains\footnote{This simplified Gaussian example is used for illustration purpose.}.
The correlation between domain-specific factors should be captured, even when they are approximated within disparate domains.
Moreover, the marginal distributions $p(\boldsymbol{U}_x)$ and $p(\boldsymbol{U}_y)$, due to coming from the same user~\cite{cao2022disencdr, huang2020causal}, should match a unique joint distribution, namely \textit{joint identifiability}.
None of the previous methods, however, are tailored to confirm the identifiability of $P(\boldsymbol{X}, \boldsymbol{Y})$, rendering them unable to uniquely recover latent factors and vulnerable to distribution shifts between domains~\cite{wang2019transfer, chen2022generative}.

Accordingly, we propose to model a cross-domain user representation that uniquely derives the {\it joint identifiability} of $p(\boldsymbol{U}_{x}, \boldsymbol{U}_{y})$.
This allows us to disentangle the user representation into a shared portion for cross-domain consistent behavioral preference, and an adaptable portion for domain-specific patterns~\cite{huang2020causal, zhang2020domain, von2021self}, leading to improved model generalization and interpretability.
Specifically, we incorporate two fundamental principles: {\it Feature Hierarchy}~(FH)~\cite{zeiler2014visualizing} of Deep Neural Network and {\it Minimal Change}~(MC)~\cite{brewka1993things, pagnucco2001causality} of Causal Inference as the cornerstones.
Using a Dual-stream Deep Adaptation Network-like architecture~\cite{long2015learning},  we follow the FH principle that encodes general, less domain-oriented features from the user-item interactions in respective domain with initial~(shallow) layers, whereas capturing more domain-oriented features within deeper layers~\cite{kundu2022subsidiary}.
The FH principle not only bridges the distributional discrepancy between domains, but also leaves exploration of domain-specific information to the deeper level feature space.
Furthermore, the MC principle~\cite{khemlani2013cognitive, khemlani2015domino} praises the minimal alterations to an existing model during inference to mitigate interference.
Leveraging the MC principle and the causal relationships in user representations, we attempt to characterize the domain shifts between $X$ and $Y$ with $\boldsymbol{U}_{x}$ and $\boldsymbol{U}_{y}$.
Recall that we have considered deeper-level network activation as a comprehensive description of user preference that preserves consistency (through a shared, cross-domain stable facet) and highlights specificity (through the adaptable, cross-domain variant facet).
In line with the MC principle, we attribute the domain shifts, signified by divergences in $\boldsymbol{U}_{x}, \boldsymbol{U}_{y}$, to this variant facet, provided that {\it joint identifiability} enables the disentanglement of user representation.

More concretely, we introduce \textbf{HJID} for CDR targeting the goals outlined above.
Using the FH principle, we first divide the user representation into shallow and deeper subspaces corresponding to neural network layers.
Within the shallow network layers, we align the activations (representations) between source and target domains with Maximum Mean Discrepancy to achieve cross-domain consistency in general features.
Following, we formulate a data generation graph for deep-level representation to depict the distribution shifts within CDR.
In this graph, we disentangle deeper layer representations into cross-domain stable and variant latent factors, defining their causal relationship using a latent variable generative model.
According to the MC principle, the stable portion remains consistent across domains, ensuring the continuity of intrinsic user preference.
On the other hand, we enforce the variant part to model cross-domain user variances under tractable and invertible transformations.
We implement a generative Flow model~\cite{chen2018neural} to map the domain-specific factors between the source and the target domain, which characterizes how these factors in one domain relate to user behaviors in another.
This allows us to reduce the impact of distribution shifts while ensuring identifiability of the joint distribution over cross-domain user representations.
Notably, our approach focuses on modeling the distributional trends with randomly sampled groups of users, sidestepping the need for exact overlap of user behaviors in two domains.
That is, it is universally applicable to CDR tasks, irrespective of whether users overlap.

The contributions of our work are summarized as follows:
\begin{itemize}
    \item For overall motivation, we present \textbf{HJID}, a hierarchical, generative, and causal discovery-driven approach for {\it identifiable} joint distribution of cross-domain user representation, through disentangling user representation into cross-domain stable and variant components. 
    Namely, \textbf{HJID} highlights {\it joint identifiability} in CDR, enabling unique latent factor recovery with limited user-item interactions and thus improved robustness under uncertain CDR scenarios.
    \item For disentanglement, we formulate a causal data generation graph to identify cross-domain stable facet from variant facet, referred to as domain-shared and domain-specific latent factors.
    \item For domain shift, we challenge the previous assumption on mutually exclusive domain-specific factors, instead learning to capture their correlations with invertible transformations, ensuring {\it joint identifiability} under distribution shift.
\end{itemize}

\begin{figure}[t]
    \centering
    \includegraphics[width=\linewidth]{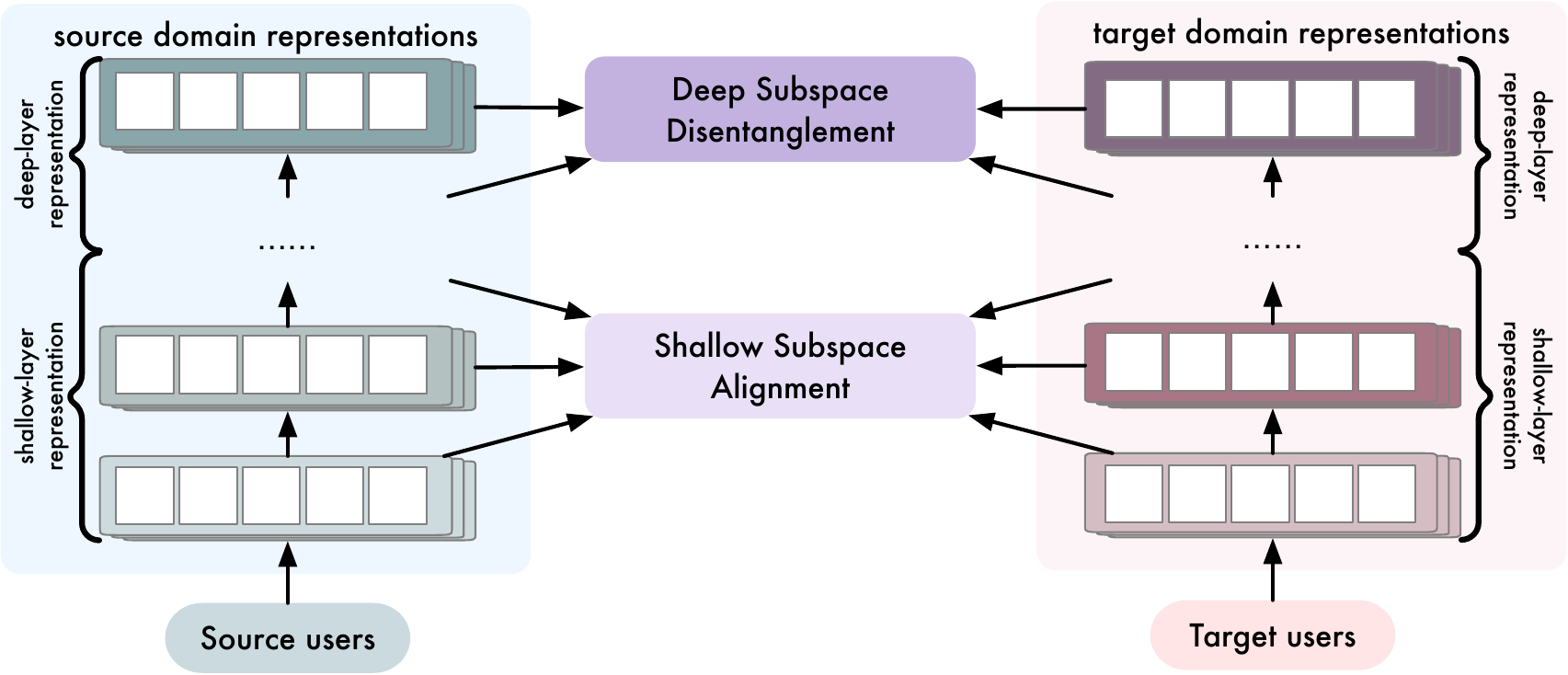}
    \caption{Overview of HJID. Following the FH principle, it separates the user representations into shallow and deep subspaces corresponding to the neural encoder layers. The shallow subspace encodes general features, enforced to be aligned between domains; while the deep subspace parameterizes more domain-oriented factors which are further disentangled into stable domain-shared and variant domain-specific facet, identified by a causal discovery structure.}
    \label{fig:overview}
    \vspace{-5mm}
\end{figure}

\section{Preliminaries}
Let $\{X, Y \}$ be two recommendation domains.
In domain $X$, we have a user set $\mathbb{U}_{x}=\{ u_i\}^{|\mathbb{U}_{x}|}$ and an item set $\mathbb{V}_{x}=\{ v_j \}^{|\mathbb{V}_{x}|}$.
The interaction set $\boldsymbol{E}_{x}= \langle u_i, v_j \rangle, \, \forall u_i \in \mathbb{U}_{x}, \forall v_j \in \mathbb{V}_{x}$ with $ \langle u_i, v_j \rangle =1$ indicating user $i$ has interacted with item $j$, else $ \langle u_i, v_j \rangle =0$.
Similarly, we have $\mathbb{U}_{y}, \mathbb{V}_{y}, \mathbb{E}_{y}$ refer to user, item, and interaction set for the domain $\boldsymbol{Y}$, respectively.
Given $\mathbb{E}_{x}, \mathbb{E}_{y}$ as training data, and new query users {\it with no} interactions observed on the target domain, our objective is to predict the next item of interaction.
We study a bi-directional CDR setting following~\cite{cao2022cross, du2023distributional}, where either $X$ or $Y$ can serve as the source domain for predictions in the other.

To facilitate this CDR prediction, we ensure the identifiability of the joint distribution over user representations across domains.

\begin{definition}[\textbf{Joint Identifiability}]
    Consider the set of latent factors $\mathcal{Z}$ and the mapping function $f: \boldsymbol{Z} \rightarrow P(\boldsymbol{U}_x, \boldsymbol{U}_y | \boldsymbol{Z})$, with joint distribution $P(\boldsymbol{U}_x, \boldsymbol{U}_y | \boldsymbol{Z})$ denotes the cross-domain user representation.
    {\it Joint identifiability} in contexts of CDR is achieved if and only if $f$ is injective,
\begin{equation}
    \boldsymbol{Z}_1 \neq \boldsymbol{Z}_2 \iff P(\boldsymbol{U}_x, \boldsymbol{U}_y | \boldsymbol{Z}_1) \neq P(\boldsymbol{U}_x, \boldsymbol{U}_y | \boldsymbol{Z}_2).
\end{equation}
$\boldsymbol{Z} \in \mathcal{Z}$ uniquely determines a joint distribution, establishing {\it joint identifiability} across domains $\boldsymbol{X}$ and $\boldsymbol{Y}$.
\end{definition}

\begin{definition}[\textbf{Marginal Identifiability}]
    In the CDR context, {\it joint identifiability} is relaxed to {\it marginal identifiability} if, there exist both marginal distributions $P(\boldsymbol{U}_x | \boldsymbol{Z}_1)$ and $P(\boldsymbol{U}_y | \boldsymbol{Z}_2)$, individually determined by latent factors $\boldsymbol{Z}_1, \boldsymbol{Z}_2 \in \mathcal{Z}$.
   For mapping function $f: \boldsymbol{Z} \rightarrow P(\boldsymbol{U}_x | \boldsymbol{Z})$ and $f: \boldsymbol{Z} \rightarrow P(\boldsymbol{U}_y | \boldsymbol{Z})$, such that
\begin{equation}
    \boldsymbol{Z}_1 \neq \boldsymbol{Z}_2 \iff P(\boldsymbol{U}_x | \boldsymbol{Z}_1) \neq P(\boldsymbol{U}_x | \boldsymbol{Z}_2) \iff P(\boldsymbol{U}_y | \boldsymbol{Z}_1) \neq P(\boldsymbol{U}_y | \boldsymbol{Z}_2)
\end{equation}
establishes the marginal identifiability within each domain.
In relation to {\it joint identifiability}, {\it marginal identifiability} does not ensure the latent factor uniqueness across the cross-domain joint space.
\end{definition}


\section{Methodology}
Fig. (\ref{fig:overview}) conceptually overviews the proposed HJID.
We begin by generating user and item representations for both domains using stacked Graph Neural Network layers~\cite{cao2022disencdr}.
Following the FH principle, we decompose user representations into two subspaces, characterized by shallow-layer and deep-layer representations.
For the former, we employ Maximum Mean Discrepancy~(MMD) that aligns the low-level, generic representations across domains and enforces {\it Shallow Subspace Alignment}.
For the latter, we introduce {\it Deep Subspace Disentanglement} anchored in the MC principle, to disentangle the behavioral commonality and variations of user representations across domains.
To achieve such disentanglement, we build a causal data generation graph that identifies the cross-domain shared latent factors from cross-domain variant ones based on {\it identifiability} of the cross-domain joint distribution.
In an effort to characterize the behavioral variance caused by domain shifts,
we further employ a generative Flow model~\cite{chen2018neural} to approximate the transformation of the variant factors between domains, ensuring {\it joint identifiability} for unique parameter recovery.

\subsection{Encoding Representations}
Previously, CDR methods often relied on strong cross-domain correlations and the concurrence of both shared and specific information among domains.
Yet, their performances degrade as domain correlations weaken. 
We thus adopt the Feature Hierarchy~(FH) principle~\cite{zeiler2014visualizing, long2015learning} to initially decouple intertwined user representations.

\subsubsection{Stackable Representation}
Denote randomly initialized embedding for each user and item by $\boldsymbol{u}_{D}$ and $\boldsymbol{v}_{D}$ with $D=\{ x, y\}$ being the domain identifier. 
Our model is architecture-agnostic, thus any neural network like MLP~\cite{raiko2012deep}, CNN~\cite{lecun1998gradient}, GCN~\cite{hamilton2017inductive} can be applicable.
To demonstrate, we introduce two $K$-layer VBGEs~\cite{cao2022disencdr} to produce the representations of users and items in both domains.
\begin{equation}
    \boldsymbol{U}_{D}^{k} \leftarrow \left\{ 
    \begin{aligned}
        & VBGE_k(\overline{\boldsymbol{A}}_{D}, \boldsymbol{u}_{D}), k=1 \\
        & VBGE_k(\overline{\boldsymbol{A}}_{D}, \boldsymbol{U}_{D}^{k-1}), k \geq 2
    \end{aligned}
    \right.
    \boldsymbol{V}_{x}^{D} \leftarrow \left\{ 
    \begin{aligned}
        & VBGE_k(\overline{\boldsymbol{A}}^{\top}_{D}, \boldsymbol{v}_{D}), k=1 \\
        & VBGE_k(\overline{\boldsymbol{A}}^{\top}_{D}, \boldsymbol{V}_{D}^{k-1}), k \geq 2
    \end{aligned}
    \right.
\end{equation}
where $\overline{\boldsymbol{A}}_{D}$ is the normalized matrix form of interaction sets $\boldsymbol{E}_{D}(D=\{ x, y\})$ and $\overline{\boldsymbol{A}}^\top_{D}$ is the transpose.
Since this is not the focus of this study, we omit the details here and refer interested readers to~\cite{cao2022disencdr, du2023distributional}.

\subsubsection{Decoupled User Representations}
Building upon the FH principle~\cite{zeiler2014visualizing}, we consider shallow subspace to encode universally shared features and deep subspace focuses on more abstract and domain-oriented features.
Our hypothesis posits that the general, domain-irrelevant features are encoded in the initial $k$ layers, whereas the subsequent $K-k$ layers specialize in extracting domain-oriented features.
Provided this decoupling, we obtain the users representations $\boldsymbol{U}_{x}$ by concatenating the shallow features $\boldsymbol{S}_{x}$ with deeper features $\boldsymbol{D}_{x}$,
\begin{equation}
    \begin{aligned}
        \boldsymbol{U}_x = [\boldsymbol{S}_x \| \boldsymbol{D}_x], 
        \boldsymbol{S}_x = 
        \left[ \boldsymbol{U}_{x}^{1} \| \cdots \| \boldsymbol{U}_{x}^k \right]
        &, \boldsymbol{D}_x = 
        \left[ \boldsymbol{U}_{x}^{k+1} \| \cdots \| \boldsymbol{U}_{x}^K \right] \\
        \boldsymbol{U}_y = [\boldsymbol{S}_y \| \boldsymbol{D}_y],
        \boldsymbol{S}_y = 
        \left[ \boldsymbol{U}_{y}^1 \| \cdots \| \boldsymbol{U}_{y}^k \right]
        &, \boldsymbol{D}_y = 
        \left[ \boldsymbol{U}_{y}^{k+1} \| \cdots \| \boldsymbol{U}_{y}^K \right]
    \end{aligned}
\end{equation}
with $\|$ denotes vector concatenation.
Analogously, the same operations are performed for domain $Y$ as well.

\subsection{Shallow Subspace Alignment}
The shallow layers encode universal features across domains.
To align these features in the shared shallow subspace, we adopt MMD~\cite{borgwardt2006integrating} for minimizing distributional divergence as in established domain adaptation strategies~\cite{long2015learning, long2017deep, yan2017mind}.
Specifically, we randomly sample two sets of shallow layer user representations $\boldsymbol{G}_{x} \sim \boldsymbol{S}_{x}$ and $\boldsymbol{G}_{y} \sim \boldsymbol{S}_{y}$, using Gaussian kernel $k(\cdot, \cdot)$ as the distance measure with MMD,
\begin{equation}
    \begin{aligned}
        k(\boldsymbol{s}_{x_i}, \boldsymbol{s}_{y_j})=\exp \left(-\frac{\|\boldsymbol{s}_{x_i}-\boldsymbol{s}_{y_j}\|^2}{2 \sigma^2}\right),
    \end{aligned}
\end{equation}
where $\sigma$ is the hyperparameter for the Gaussian kernel.
$\boldsymbol{s}_{x_i}, \boldsymbol{s}_{y_j}$ refers to the shallow layer representations of user $x_i$ and user $y_j$ from the sets $\boldsymbol{G}_{x}$ and $\boldsymbol{G}_{y}$.
The computed MMD over all sampled users is
\begin{equation}
    \begin{aligned}
        & MMD^2(\boldsymbol{G}_{X}, \boldsymbol{G}_{Y}) = \\
        & h \left(
        \sum_{i=1}^{N} \sum_{j=1}^{N} k(\boldsymbol{s}_{x_i}, \boldsymbol{s}_{x_j}) - 
        \sum_{i=1}^{N} \sum_{j=1}^{N} k(\boldsymbol{s}_{x_i}, \boldsymbol{s}_{y_j})
        + \sum_{i=1}^{N} \sum_{j=1}^{N} k(\boldsymbol{s}_{y_i}, \boldsymbol{s}_{y_j}) \right),
    \end{aligned}
\end{equation}
where $h = \frac{1}{N(N-1)}$ and $N$ is the sample size.
Notably, pairwise overlapping among sampled users is not a prerequisite, meaning that the alignment can also be adapted to work with CDR in the absence of overlap users.
Minimizing MMD allows us to empirically substantiate the alignment efficacy of general features, ensuring domain consistency at a shallow level.

\subsection{Deep Subspace Disentanglement}
As aforementioned, we seek a disentanglement of deeper, domain-oriented representations to be domain-shared~(stable) and domain-specific~(variant) factors,
with the former reflecting consistent user behavior and the latter accommodating distribution shifts across domains.
In general, it is difficult due to the intertwined effect of complex user behavior~\cite{cao2022cross, cao2022contrastive, du2023distributional}.
To tackle this, we present the data generation graph for these domain-oriented representations with causality and delve into how to leverage the latent variable model to ensure {\it joint identifiability}.

\begin{figure}
    \centering
    \includegraphics[width=0.9\linewidth]{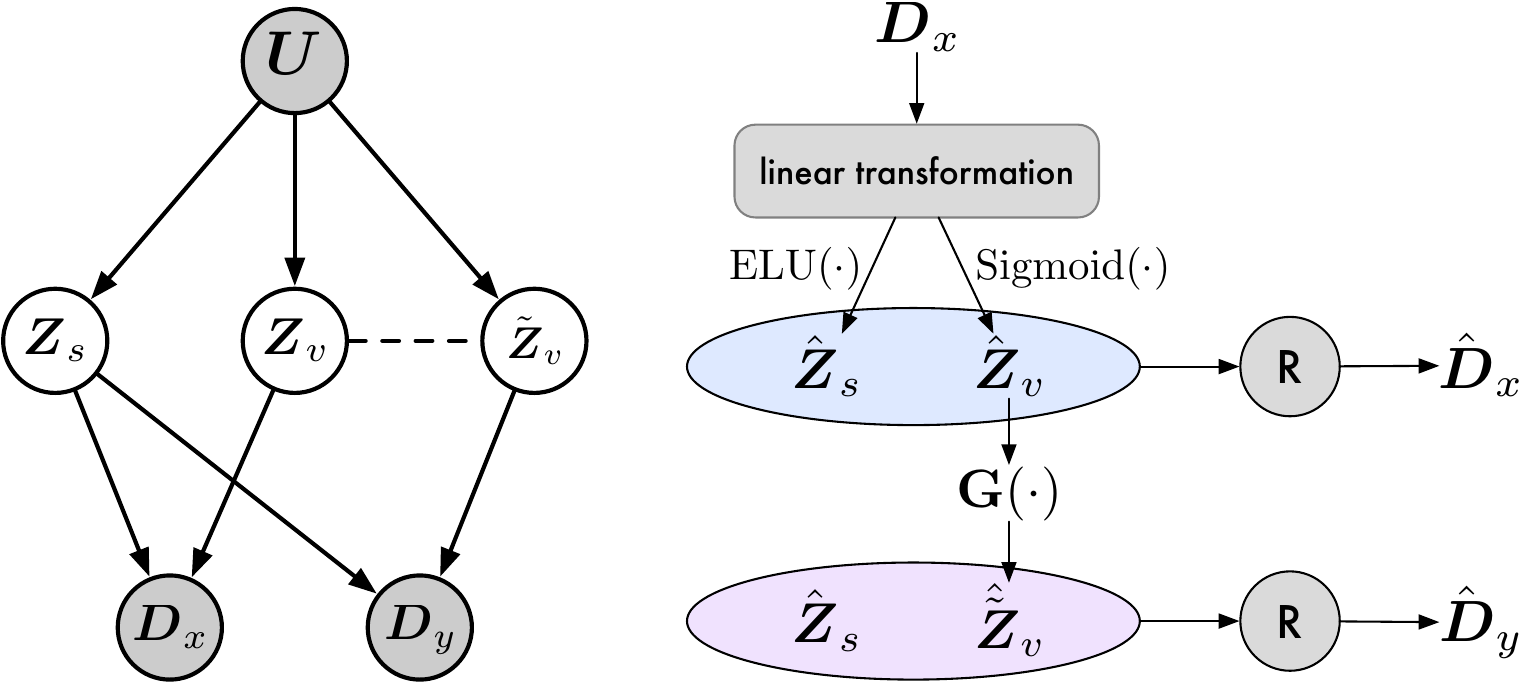}
    \caption{Causal data generation graph~(left) and framework~(right) of deep subspace disentanglement. The grey circle refers to the observed variables. The dashed line indicates a correlation between two variables. R denotes the reparameterization trick.}
    \label{fig:deep}
    \vspace{-5mm}
\end{figure}

\subsubsection{Generation Graph for CDR} 
From a probabilistic perspective, we assume the domain-oriented representations of source domain $\boldsymbol{D}_x$ are generated by latent variables $f: \boldsymbol{Z} \rightarrow \boldsymbol{D}_x$.
For disentanglement, the latent variable $\boldsymbol{Z}$ is partitioned into a) the stable portion $\boldsymbol{Z}_s$ and b) variant portion $\boldsymbol{Z}_v$.
The stable portion $\boldsymbol{Z}_s$ characterizes the consistency of the user representation distribution that remains invariant.
In contrast, $\boldsymbol{Z}_v$ addresses domain-specific nuances by capturing the observation variances.
Similarly, we have $f': \boldsymbol{Z}' \rightarrow \boldsymbol{D}_y$ for the target domain.
$\boldsymbol{Z}'$ shares the stable portion with $\boldsymbol{Z}_s$, while focusing only on the transformations between their variant portions $\tilde{\boldsymbol{Z}}_v = \mathbf{G}(\boldsymbol{Z}_v)$.
Intuitively, this interprets the observation variance as a result of domain shifts, under the MC principle.

\subsubsection{Latent Variable Disentanglement}
For simplicity, we model $\boldsymbol{Z}$ as a Gaussian with mean and variance parameterized by $\boldsymbol{Z}_s$ and $\boldsymbol{Z}_v$.
We first employ non-linear transformations on $\boldsymbol{D}_x$,
\begin{equation}
    \begin{aligned}
        & \boldsymbol{Z} \sim \mathcal{N}(\boldsymbol{Z}_s, \boldsymbol{Z}_v) = \mathcal{N}\left(\mathrm{ELU}(\boldsymbol{W}_s \boldsymbol{D}_x), \mathrm{Sigmoid}(\boldsymbol{W}_v \boldsymbol{D}_x) \right)
    \end{aligned}
\end{equation}
where the ELU and the Sigmoid are non-linear activation for $\boldsymbol{Z}_s$ and $\boldsymbol{Z}_v$, respectively. 
$\boldsymbol{W}_s$ and $\boldsymbol{W}_v$ are weight parameters, and the bias terms are omitted for brevity.

Note that $\tilde{\boldsymbol{Z}}_v$ can be exclusively transformed from $\boldsymbol{Z}_v$, without changing $\boldsymbol{Z}_s$~(See Fig.~\ref{fig:deep}).
To approximate the mapping $\mathbf{G}(\cdot)$, we opt for Normalizing Flow~\cite{zhong2020session, jeon2021lightmove, guo2022evolutionary} and take advantage of the bijective mappings.
Namely, this enables both forward and inverse transformations exact, which facilitates {\it joint identifiability} of $p(\boldsymbol{U}_x, \boldsymbol{U}_y)$ by ensuring $\boldsymbol{Z}_{v}$ maps uniquely to $\tilde{\boldsymbol{Z}}_v$, and vice versa.
Consider the marginal distribution $P_{X}(\boldsymbol{Z}_v)$ and $P_{Y}(\tilde{\boldsymbol{Z}}_v)$, we can derive the density function of $\tilde{\boldsymbol{Z}}_v$ by
\begin{equation}
\label{eq:flow}
P_{Y}(\tilde{\boldsymbol{Z}}_v) =P_{X}(\mathbf{f}(\tilde{\boldsymbol{Z}}_v))\operatorname{det}(\operatorname{Df}(\tilde{\boldsymbol{Z}}_v)) =P_{X}(\boldsymbol{Z}_v) \operatorname{det} (\operatorname{Dg}(\boldsymbol{Z}_v))^{-1}
\end{equation}
where $\mathbf{f}=\mathbf{g}^{-1}$ is the inverse of $\mathbf{g}$. $\operatorname{Df}(\cdot)=\frac{\partial \mathbf{f}}{\partial \cdot}$ is the Jacobian of $\mathbf{f}$ and $\operatorname{Dg}(\cdot)=\frac{\partial \mathbf{g}}{\partial \cdot}$ is the Jacobian of $\mathbf{g}$.
$\text{det}(\cdot)$ is the matrix determinant.
Usually, the limited model expressiveness comes as the downside of bijective transformation.
We thus chain a $L$-layer bijections $\mathbf{G}=\mathbf{g}_1 \circ \cdots \circ \mathbf{g}_L$ to improve the model flexibility for when fitting non-linear patterns.
Equivalently, Eq.(\ref{eq:flow}) now becomes:
\begin{equation}
    \begin{aligned}
        P_Y(\tilde{\boldsymbol{Z}}_v)
        = P_X( \mathbf{G}(\boldsymbol{Z}_v))
        = P_X(\boldsymbol{Z}_v) \cdot \prod_{l=0}^{L} \operatorname{det} (\operatorname{Dg_l}\left(\boldsymbol{Z}_{l}\right))^{-1},
    \end{aligned}
\end{equation}
with
\begin{equation}
   \begin{aligned}
    \operatorname{Dg}(\boldsymbol{Z}_v) 
    =\left[\frac{\partial \boldsymbol{Z}_1}{\partial \boldsymbol{Z}_0}\right]\left[\frac{\partial \boldsymbol{Z}_2}{\partial \boldsymbol{Z}_1}\right] \cdots\left[\frac{\partial \boldsymbol{Z}_v}{\partial \boldsymbol{Z}_{L-1}}\right]
    =\prod_{l=0}^{L} \operatorname{Dg_l}\left(\boldsymbol{Z}_{l}\right),
    \end{aligned}
\end{equation}
The implementation of transformation $\operatorname{Dg_i}(\cdot)$ can be of any choice~\cite{papamakarios2017masked, huang2018neural, chen2018neural, rezende2020normalizing}.
We further use the reparameterization trick~\cite{kingma2013auto} to sample the domain-oriented representations $\hat{\boldsymbol{D}}_x, \hat{\boldsymbol{D}}_y$ refined by the estimated latent factors $\hat{\boldsymbol{Z}}_s$, $\hat{\boldsymbol{Z}}_v$ and $\hat{\tilde{\boldsymbol{Z}}}_v$,
\begin{equation}
    \begin{aligned}
        \hat{\boldsymbol{D}}_x = \boldsymbol{Z}_s + \boldsymbol{Z}_v \odot \mathbf{\epsilon}_x, 
        \hat{\boldsymbol{D}}_y = \boldsymbol{Z}_s + \tilde{\boldsymbol{Z}}_v \odot \mathbf{\epsilon}_y,
    \end{aligned}
\end{equation}
with $\mathbf{\epsilon}_x \sim N(\mathbf{0}, \mathbf{I})$, 
$\mathbf{\epsilon}_y \sim N(\mathbf{0}, \mathbf{I})$, and 
$\tilde{\boldsymbol{Z}}_v=\mathbf{G}(\boldsymbol{Z}_v)$.
$\odot$ denotes element-wise multiplication.
We substitute $\boldsymbol{D}_x, \boldsymbol{D}_y$ with $\hat{\boldsymbol{D}}_x, \hat{\boldsymbol{D}}_y$, leading to refined user representations $\hat{\boldsymbol{U}}_x = [\boldsymbol{S}_x \| \hat{\boldsymbol{D}}_x]$ and $\hat{\boldsymbol{U}}_y = [\boldsymbol{S}_y \| \hat{\boldsymbol{D}}_y]$.

\subsubsection{Marginal vs. Joint Identifiability}
We note that what existing methods at most achieve is {\it marginal identifiability}, as they primarily strive to maximize marginal likelihood $P(\boldsymbol{U}_x)$ or $P(\boldsymbol{U}_y)$, ``averaging out'' possible influences of other domains.
However, this marginalization loses the conditionality between $\boldsymbol{U}_x$ and $\boldsymbol{U}_y$ captured by their joint distribution.
Instead, our method leverages Flow-based invertible transformations to ensure that $P(\boldsymbol{U}_x, \boldsymbol{U}_y)$ can be uniquely recovered by $\boldsymbol{Z}_{v}$ and $\tilde{\boldsymbol{Z}}_{v}$ jointly.

\subsection{Optimization}
Having obtained shallow feature consistency and deeper feature disentanglement, we now describe the multi-task objective functions for HJID.
Given random groups of users $\boldsymbol{G}_X, \boldsymbol{G}_Y$, the objectives include 
1) MMD that minimizes the divergence between the source and the target shallow representations,
\begin{equation}
    \mathcal{L}_s = MMD^2(\boldsymbol{G}_X, \boldsymbol{G}_Y).
\end{equation}
2) The deeper subspace disentanglement involves a Flow-based model
$\mathbf{G}(\cdot)$, which maximizes the predictive likelihood on the target domain after a chain of transformations $\boldsymbol{G}(\boldsymbol{Z}_v)$,
\begin{equation}
    \begin{aligned}
    \mathcal{L}_G =\sum_{i=1}^N  \left[ \log p_{Y} \left( \boldsymbol{G} (\mathbf{Z}_v^{(i)}) \right) - \sum_{l=0}^{L} \log  \operatorname{det} \left( \operatorname{Dg_l}\left(\mathbf{Z}_v^{(i)}\right) \right) \right].
    \end{aligned}
\end{equation}
3) To perform accurate CDR predictions, we reconstruct user-item interactions using the inferred latent variables $\boldsymbol{Z}$.
This objective serves dual purposes: stabilizing the optimization of generative model $\boldsymbol{G}(\cdot)$ and enforcing the causal relation between CDR prediction and $\boldsymbol{Z}$~\cite{kong2022partial}.
Moreover, we leverage variational information bottleneck~(VIB)~\cite{alemi2016deep} to constrain the optimization.
Being renowned for excelling at producing informative representations~\cite{cao2022disencdr, du2023distributional}, VIB maximizes the mutual information between $\boldsymbol{Z}$ and refined representations $\hat{\boldsymbol{D}}_x$ and $\hat{\boldsymbol{D}}_y$ and removes the redundancy therefrom.
\cite{du2023distributional} shows that the empirical lower bound of VIB can be approximated by the binary cross-entropy loss as
\begin{equation}
    \begin{aligned}
        & I((\boldsymbol{Z}_s, \boldsymbol{Z}_v); \hat{\boldsymbol{D}}_x) \geq       \\
        & \sum_{ \langle u_i,v_j \rangle \in \mathbb{E}_X} \log \sigma(\boldsymbol{V}_x, \hat{\boldsymbol{D}}_x) + \sum_{\langle u_i,v_j \rangle \notin \mathbb{E}_X} \log (1-\sigma(\boldsymbol{V}_x, \hat{\boldsymbol{D}}_x)), \\
        & I((\boldsymbol{Z}_s, \tilde{\boldsymbol{Z}}_v); \hat{\boldsymbol{D}}_y) \geq \\
        & \sum_{ \langle u_i,v_j \rangle \in \mathbb{E}_Y} \log \sigma(\boldsymbol{V}_y, \hat{\boldsymbol{D}}_y) + \sum_{\langle u_i,v_j \rangle \notin \mathbb{E}_Y} \log (1-\sigma (\boldsymbol{V}_y, \hat{\boldsymbol{D}}_y)), 
    \end{aligned}
\end{equation}
where $\sigma(\cdot)$ is inner product with Sigmoid.

Summing all tasks up yields the overall optimization objective,
\begin{equation}
    \begin{aligned}
        \mathcal{L} = & \mathcal{L}_s + \mathcal{L}_G + I((\boldsymbol{Z}_s, \boldsymbol{Z}_v); \hat{\boldsymbol{D}}_x) + I((\boldsymbol{Z}_s, \tilde{\boldsymbol{Z}}_v); \hat{\boldsymbol{D}}_y).
    \end{aligned}
\end{equation}

\section{Experiments}
Here are the research questions we aim to answer:
\begin{itemize}
    \item[RQ1)]
    Does HJID outperform existing CDR methods?
    \item[RQ2)]
    If needed, does HJID still demonstrate efficacy even with challenging non-overlap CDR predictions?
    \item[RQ3)]
    Do individual components contribute positively to HJID?
    \item[RQ4)]
    How do hyperparameters impact model performance?
\end{itemize}

\subsection{Experimental setup}
\subsubsection{Datasets}
We conduct all empirical studies on the widely studied Amazon reviews datasets\footnote{https://cseweb.ucsd.edu/~jmcauley/datasets/amazon/links.html}.
To showcase the impact of domain correlations, we build 6 pairwise CDR settings: Game-Video and Cloth-Sport are strong-correlation tasks, and Game-Cloth, Game-Sport, Video-cloth, and Video-Game are weak-correlation tasks.
All the baselines and our model are evaluated on 6 tasks to evaluate the model performance comprehensively.
The detailed information is shown in Table.\ref{tab: data}.
\begin{table}
    \caption{Dataset statistics.}
    \label{tab: data}
    \centering
    \resizebox{0.8\linewidth}{!}{
    \begin{tabular}{l|c|c|c|c|c|c|c}
    \hline
    No.                     & Domain & user   & \#Overlap               & item   & interactions & Testing                & validation             \\
    \hline
    \multirow{2}{*}{Task 1} & Game   & 25,025 & \multirow{2}{*}{2,179} & 12,319 & 157,721      & \multirow{2}{*}{443}   & \multirow{2}{*}{434}   \\
                            & Video  & 19,457 &                        & 8,751  & 158,984      &                        &                        \\
    \hline
    \multirow{2}{*}{Task 2} & Cloth  & 41,829 & \multirow{2}{*}{9,828} & 17,943 & 194,121      & \multirow{2}{*}{1,972} & \multirow{2}{*}{1,964} \\
                            & Sport  & 27,328 &                        & 12,655 & 170,426      &                        &                        \\
    \hline
    \multirow{2}{*}{Task 3} & Game   & 17,299 & \multirow{2}{*}{5,968} & 9,125  & 104,034      & \multirow{2}{*}{1,200} & \multirow{2}{*}{1,192} \\
                            & Cloth  & 57,929 &                        & 25,196 & 297,241      &                        &                        \\
    \hline
    \multirow{2}{*}{Task 4} & Game   & 19,465 & \multirow{2}{*}{3,442} & 9,834  & 116,914      & \multirow{2}{*}{690}   & \multirow{2}{*}{688}   \\
                            & Sport  & 38,996 &                        & 16,964 & 255,959      &                        &                        \\
    \hline
    \multirow{2}{*}{Task 5} & Video  & 19,623 & \multirow{2}{*}{2,001} & 8,989  & 167,615      & \multirow{2}{*}{401}   & \multirow{2}{*}{400}   \\
                            & Cloth  & 70,183 &                        & 31,282 & 388,224      &                        &                        \\
    \hline
    \multirow{2}{*}{Task 6} & Video  & 19,648 & \multirow{2}{*}{2,159} & 8,781  & 164,457      & \multirow{2}{*}{439}   & \multirow{2}{*}{430}   \\
                            & Sport  & 41,939 &                        & 18,208 & 278,036      &                        &                        \\
    \hline
    \end{tabular}
    }
    \vspace{-5mm}
\end{table}

\subsubsection{Compared methods}
We introduce eight baselines for comparisons:
For \textbf{Alignment} methods, we import \textbf{CMF}~\cite{singh2008relational} and \textbf{LFM}~\cite{agarwal2011localized}.
For \textbf{Bridge} methods, we employ \textbf{EMCDR}~\cite{man2017cross}, implemented with two variants: \textbf{EMCDR-MF} in Matrix Factorization and \textbf{EMCDR-NGCF} in Neural Graph Collaborative Filtering\cite{wang2019neural}, and \textbf{PTUPCDR}~\cite{zhu2022personalized}.
For the latest \textbf{Constraint} baselines, four methods are introduced: \textbf{DisenCDR}~\cite{cao2022disencdr}, \textbf{UniCDR}~\cite{cao2023towards}, \textbf{CDRIB}~\cite{cao2022cross}, and \textbf{DPMCDR}~\cite{du2023distributional}.

\subsubsection{Implementation details}
We divide all overlapped users into training(60\%), testing(20\%), and validation(20\%). 
All non-overlapped users are only used for training.
Following common practice~\cite{krichene2020sampled} as with most studies, we evaluate our model and comparison methods using {\it leave-one-out} method: only 1 positive item and 999 random-selected negative items are calculated.
We report 3 evaluation metrics of predictive performance: Mean Reciprocal Rank~(MRR), Hit Rate~(HR@\{10, 20, 30\}), and Normalized Discounted Cumulative Gain~(NDCG@\{10, 20, 30\}).
The implementations of \textbf{CMF}, \textbf{LFM}, \textbf{EMCDR}, \textbf{PTUPCDR}, \textbf{DisenCDR},  \textbf{UniCDR}, and \textbf{CDRIB} follow the open-sourced codebases, implemented with PyTorch.
We implement \textbf{DPMCDR} upon CDRIB and make modifications according to~\cite{du2023distributional}.
The following are the default settings.
The number of VBGE layers $K$ is set to 3.
We set the number of shallow layers $k$ to 2.
That is, shallow representations and deeper representations become $\boldsymbol{S}_x=[ \boldsymbol{U}_x^1 \| \boldsymbol{U}_x^2]$ and $\boldsymbol{D}_x=[\boldsymbol{U}_x^3]$, ($\boldsymbol{S}_y=[ \boldsymbol{U}_y^1 \| \boldsymbol{U}_y^2]$ and $\boldsymbol{D}_y=[\boldsymbol{U}_y^3]$ for the target domain).
We choose ELU as activation for $\boldsymbol{Z}_s$ for its smooth saturation than the standard ReLU~\cite{grelsson2018improved} and Sigmoid activation for $\boldsymbol{Z}_v$ since its sensitivity to different distributions~\cite{zhang2016dnn}.
We search embedding size $d$ in the range of \{8, 16, 32, 64, 128\}. 
The size of user group $N$ is searched within \{16, 32, 64, 128, 256, 512, 1024\}. 
To determine how specific Flows affect the performance, 
we include 4 transformations~MAF\cite{papamakarios2017masked}, NAF\cite{huang2018neural}, NODE\cite{chen2018neural}, and NCSF\cite{rezende2020normalizing}, according to the open-sourced code\footnote{https://github.com/francois-rozet/zuko}.
Length of the invertible bijection chain $L$ is searched over \{1, 2, 3, 4, 5\}. 
We also investigate the influence of different shallow depths $k$=\{1, 2, 3, 4\}.
We use Adam for model optimization.


\begin{table*}
    \caption{Overall performance(\%).
    Improved shows the improvement over the runner-up results.
    * indicates that the improvement is statistically significant when the HJID and the best baseline are compared with a paired t-test level of p$<0.05$.}
    \label{tab: overall}
    \vspace{-2mm}
    \centering
    \resizebox{\linewidth}{!}{
    \begin{tabular}{l|ccccccc|ccccccc}
    \hline
    \multicolumn{1}{c|}{\multirow{2}{*}{Methods}} & \multicolumn{7}{c|}{Game}                                                                                                                                        & \multicolumn{7}{c}{Video}                                                                                                                                        \\ \cline{2-15} 
    \multicolumn{1}{c|}{}                         & MRR                  & NDCG@10              & NDCG@20              & NDCG@30              & HR@10                & HR@20                 & HR@30                 & MRR                  & NDCG@10              & NDCG@20              & NDCG@30              & HR@10                & HR@20                 & HR@30                 \\ \hline
    CMF                                           & 0.91($\pm$0.08)          & 0.41($\pm$0.15)          & 0.61($\pm$0.10)          & 1.05($\pm$0.08)          & 1.04($\pm$0.31)          & 1.82($\pm$0.32)           & 3.91($\pm$0.55)           & 0.84($\pm$0.07)          & 0.45($\pm$0.08)          & 0.82($\pm$0.08)          & 0.98($\pm$0.08)          & 1.30($\pm$0.10)          & 2.86($\pm$0.14)           & 3.65($\pm$0.80)           \\
    LFM                                           & 1.68(±0.13)              & 1.18(±0.13)              & 1.54(±0.10)               & 1.9(±0.12)              & 2.44(±0.16)              & 3.84(±0.16)               & 5.6(±0.37)                & 1.86(±0.24)              & 1.36(±0.22)              & 1.82(±0.23)              & 2.14(±0.23)              & 2.76(±0.25)              & 4.56(±0.27)               & 6.08(±0.29) 
    \\ \hline
    EMCDR-MF                                      & 1.46($\pm$0.37)          & 0.89($\pm$0.35)          & 1.46($\pm$0.36)          & 1.80($\pm$0.41)          & 1.56($\pm$0.55)          & 3.91($\pm$0.55)           & 5.47($\pm$0.55)           & 0.98($\pm$0.55)          & 0.64($\pm$0.13)          & 0.97($\pm$0.07)          & 1.25($\pm$0.39)          & 1.82($\pm$0.38)          & 3.13($\pm$0.32)           & 4.43($\pm$0.39)           \\
    EMCDR-NGCF                                    & 1.51($\pm$0.08)          & 1.07($\pm$0.23)          & 1.65($\pm$0.10)          & 2.26($\pm$0.34)          & 2.34($\pm$1.10)          & 4.69($\pm$0.55)           & 7.55($\pm$1.69)           & 1.11($\pm$0.11)          & 0.70($\pm$0.27)          & 0.98($\pm$0.46)          & 1.14($\pm$0.36)          & 1.56($\pm$0.84)          & 2.86($\pm$0.66)           & 3.65($\pm$0.85)           \\
    PTUPCDR                                       & 1.76($\pm$0.45)          & 1.23($\pm$0.69)          & 2.00($\pm$0.55)          & 2.61($\pm$0.43)          & 2.34($\pm$1.46)          & 5.47($\pm$0.96)           & 8.33($\pm$0.32)           & 1.39($\pm$0.28)          & 0.89($\pm$0.19)          & 1.34($\pm$0.39)          & 1.72($\pm$0.27)          & 1.56($\pm$0.46)          & 3.39($\pm$0.59)           & 5.73($\pm$0.76)           \\ \hline
    DisenCDR                                      & 2.15($\pm$0.10)          & 1.52($\pm$0.21)          & 2.32($\pm$0.22)          & 3.31($\pm$0.37)          & 3.20($\pm$0.60)          & 6.41($\pm$0.81)           & 11.06($\pm$1.41)          & 2.26($\pm$0.31)          & 1.59($\pm$0.41)          & 2.81($\pm$0.52)          & 3.79($\pm$0.54)          & 3.57($\pm$0.89)          & 8.43($\pm$1.40)           & 13.02($\pm$1.45)          \\
    UniCDR                                        & 3.02($\pm$0.10)          & 2.93($\pm$0.21)          & 3.74($\pm$0.22)          & 4.21($\pm$0.34)          & 5.05($\pm$0.70)          & 8.63($\pm$0.47)           & 11.18($\pm$0.74)          & 2.51($\pm$0.38)          & 2.56($\pm$0.30)          & 3.23($\pm$0.26)          & 3.81($\pm$0.28)          & 5.14($\pm$0.44)          & 7.80($\pm$0.62)           & 10.51($\pm$0.72)          \\
    CDRIB                                         & 4.01($\pm$0.23)          & 4.52($\pm$0.27)          & 5.80($\pm$0.35)          & 6.65($\pm$0.28)          & 9.32($\pm$0.41)          & 14.65($\pm$0.74)          & 18.64($\pm$0.38)          & {\ul 3.78($\pm$0.46)}    & {\ul 3.84($\pm$0.51)}    & {\ul 5.22($\pm$0.51)}    & {\ul 6.12($\pm$0.64)}    & 7.60($\pm$0.74)          & {\ul 13.09($\pm$0.73)}    & {\ul 17.33($\pm$1.36)}    \\
    DPMCDR                                        & {\ul 4.45($\pm$0.36)}    & {\ul 4.80($\pm$0.41)}    & {\ul 6.27($\pm$0.28)}    & {\ul 7.14($\pm$0.36)}    & {\ul 9.51($\pm$0.67)}    & {\ul 15.22($\pm$0.57)}    & {\ul 19.31($\pm$0.72)}    & 3.67($\pm$0.14)          & 3.83($\pm$0.14)          & 5.06($\pm$0.30)          & 5.96($\pm$0.30)          & {\ul 7.89($\pm$0.32)}    & 12.80($\pm$0.93)          & 17.04($\pm$1.00)          \\ \hline
    \textbf{HJID*}                                          & \textbf{4.63($\pm$0.10)*} & \textbf{4.97($\pm$0.06)*} & \textbf{6.57($\pm$0.10)*} & \textbf{7.43($\pm$0.11)*} & \textbf{9.61($\pm$0.21)*} & \textbf{16.07($\pm$0.20)*} & \textbf{20.11($\pm$0.17)*} & \textbf{4.39($\pm$0.21)*} & \textbf{4.67($\pm$0.19)*} & \textbf{6.00($\pm$0.12)*} & \textbf{6.95($\pm$0.22)*} & \textbf{9.16($\pm$0.15)*} & \textbf{14.48($\pm$0.22)*} & \textbf{18.95($\pm$0.41)*} \\
    Improved                                      & 3.92\%                & 3.57\%                & 4.84\%                & 4.12\%                & 1.00\%                & 5.63\%                 & 4.19\%                 & 16.08\%               & 21.45\%               & 14.94\%               & 13.57\%               & 16.05\%               & 10.62\%                & 9.32\%                \\ \hline
    \end{tabular}
    }

    \centering
    \resizebox{\linewidth}{!}{
    \begin{tabular}{l|ccccccc|ccccccc}
    \hline
    \multicolumn{1}{c|}{\multirow{2}{*}{Methods}} & \multicolumn{7}{c|}{Cloth}                                                                                                                                       & \multicolumn{7}{c}{Sport}                                                                                                                                        \\ \cline{2-15} 
    \multicolumn{1}{c|}{}                         & MRR                  & NDCG@10              & NDCG@20              & NDCG@30              & HR@10                & HR@20                 & HR@30                 & MRR                  & NDCG@10              & NDCG@20              & NDCG@30              & HR@10                & HR@20                 & HR@30                 \\ \hline
    CMF                                           & 1.42($\pm$0.17)          & 1.06($\pm$0.15)          & 1.73($\pm$0.10)          & 2.00($\pm$0.21)          & 2.60($\pm$0.43)          & 4.95($\pm$0.64)           & 6.25($\pm$0.32)           & 1.29($\pm$0.12)          & 1.00($\pm$0.30)          & 1.59($\pm$0.23)          & 1.86($\pm$0.12)          & 2.60($\pm$0.96)          & 4.95($\pm$0.64)           & 6.25($\pm$0.32)           \\
    LFM                                           & 1.48(±0.02)              & 1.02(±0.06)              & 1.46(±0.04)              & 1.84(±0.03)              & 2.42(±0.10)              & 4.16(±0.13)               & 5.94(±0.21)               & 1.56(±0.22)              & 1.06(±0.23)              & 1.52(±0.24)              & 1.96(±0.22)              & 2.44(±0.31)              & 4.32(±0.38)               & 6.40(±0.33)
    \\ \hline
    EMCDR-MF                                      & 1.55($\pm$0.69)          & 1.27($\pm$0.11)          & 1.40($\pm$0.04)          & 2.08($\pm$0.11)          & 2.60($\pm$0.32)          & 4.17($\pm$0.32)           & 6.16($\pm$0.91)           & 2.04($\pm$0.46)          & 1.49($\pm$0.46)          & 1.94($\pm$0.38)          & 2.22($\pm$0.44)          & 2.08($\pm$0.32)          & 4.17($\pm$0.31)           & 5.21($\pm$0.32)           \\
    EMCDR-NGCF                                    & 1.98($\pm$0.69)          & 1.64($\pm$0.85)          & 1.98($\pm$0.70)          & 2.37($\pm$0.42)          & 2.34($\pm$0.96)          & 3.65($\pm$0.32)           & 5.47($\pm$0.96)           & 2.39($\pm$0.68)          & 1.92($\pm$0.73)          & 2.50($\pm$0.72)          & 2.90($\pm$0.86)          & 3.13($\pm$0.96)          & 5.47($\pm$0.96)           & 7.29($\pm$1.59)           \\
    PTUPCDR                                       & 2.09($\pm$0.30)          & 2.15($\pm$0.21)          & 2.47($\pm$0.35)          & 2.53($\pm$0.41)          & 3.47($\pm$0.32)          & 3.65($\pm$0.64)           & 6.77($\pm$0.96)           & 1.14($\pm$0.31)          & 0.55($\pm$0.29)          & 0.89($\pm$0.52)          & 1.89($\pm$0.65)          & 1.56($\pm$0.32)          & 2.86($\pm$1.28)           & 7.55($\pm$1.91)           \\ \hline
    DisenCDR                                      & 2.13($\pm$0.12)          & 1.42($\pm$0.03)          & 2.29($\pm$0.23)          & 3.11($\pm$0.52)          & 3.08($\pm$0.13)          & 6.58($\pm$0.78)           & 10.44($\pm$1.14)          & 2.58($\pm$0.18)          & 1.54($\pm$0.08)          & 2.51($\pm$0.12)          & 3.58($\pm$0.08)          & 3.22($\pm$0.07)          & 7.09($\pm$0.27)           & 12.14($\pm$0.65)          \\
    UniCDR                                        & 2.70($\pm$0.18)          & 2.70($\pm$0.23)          & 3.48($\pm$0.29)          & 4.03($\pm$0.31)          & 5.20($\pm$0.48)          & 8.31($\pm$0.74)           & 10.93($\pm$0.83)          & 1.70($\pm$0.15)          & 1.55($\pm$0.20)          & 2.11($\pm$0.23)          & 2.51($\pm$0.25)          & 3.09($\pm$0.46)          & 5.33($\pm$0.50)           & 7.20($\pm$0.57)           \\
    CDRIB                                        & {\ul 3.24($\pm$0.16)}    & {\ul 3.40($\pm$0.16)}    & {\ul 4.39($\pm$0.13)}    & {\ul 5.14($\pm$0.17)}    & {\ul 6.54($\pm$0.24)}    & {\ul 10.80($\pm$0.17)}    & {\ul 14.35($\pm$0.24)}    & {\ul 4.17($\pm$0.13)}    & {\ul 4.44($\pm$0.12)}    & {\ul 5.75($\pm$0.17)}    & {\ul 6.61($\pm$0.16)}    & {\ul 8.68($\pm$0.25)}    & {\ul 13.91($\pm$0.31)}    & {\ul 17.95($\pm$0.33)}    \\
    DPMCDR                                       & 2.86($\pm$0.02)          & 2.88($\pm$0.04)          & 3.97($\pm$0.05)          & 4.79($\pm$0.12)          & 6.00($\pm$0.23)          & 11.03($\pm$0.12)          & 14.19($\pm$0.27)          & 4.12($\pm$0.27)          & 4.27($\pm$0.27)          & 5.54($\pm$0.37)          & 6.46($\pm$0.29)          & 8.19($\pm$0.34)          & 13.27($\pm$0.48)          & 17.63($\pm$0.37)          \\ \hline

    \textbf{HJID*}                                          & \textbf{3.43($\pm$0.24)*} & \textbf{3.50($\pm$0.30)*} & \textbf{4.53($\pm$0.31)*} & \textbf{5.31($\pm$0.28)*} & \textbf{6.79($\pm$0.47)*} & \textbf{11.90($\pm$0.53)*} & \textbf{14.54($\pm$0.39)*} & \textbf{4.61($\pm$0.04)*} & \textbf{4.85($\pm$0.08)*} & \textbf{6.26($\pm$0.06)*} & \textbf{7.15($\pm$0.07)*} & \textbf{9.17($\pm$0.30)*} & \textbf{14.77($\pm$0.16)*} & \textbf{19.08($\pm$0.23)*} \\
    Improved                                      & 5.73\%                & 3.15\%                & 3.26\%                & 3.16\%                & 3.79\%                & 10.18\%                & 1.33\%                 & 10.39\%               & 9.24\%                & 8.79\%                & 8.31\%                & 5.53\%                & 6.18\%                 & 6.28\%                 \\ \hline
    \end{tabular}
    }
    
    \centering
    \resizebox{\linewidth}{!}{
    \begin{tabular}{l|ccccccc|ccccccc}
    \hline
    \multicolumn{1}{c|}{\multirow{2}{*}{Methods}} & \multicolumn{7}{c|}{Game}                                                                                                                                        & \multicolumn{7}{c}{Cloth}                                                                                                                                        \\ \cline{2-15} 
    \multicolumn{1}{c|}{}                         & MRR                  & NDCG@10              & NDCG@20              & NDCG@30              & HR@10                & HR@20                 & HR@30                 & MRR                  & NDCG@10              & NDCG@20              & NDCG@30              & HR@10                & HR@20                 & HR@30                 \\ \hline
    CMF                                           & 1.02($\pm$0.07)          & 0.74($\pm$0.07)          & 1.05($\pm$0.09)          & 1.22($\pm$0.15)          & 2.08($\pm$0.10)          & 3.39($\pm$0.16)           & 4.17($\pm$0.47)           & 0.98($\pm$0.34)          & 0.49($\pm$0.02)          & 0.96($\pm$0.04)          & 1.28($\pm$0.09)          & 1.30($\pm$0.13)          & 3.13($\pm$0.19)           & 4.95($\pm$0.18)           \\
    LFM                                           & 1.60(±0.04)              & 1.06(±0.06)              & 1.52(±0.07)              & 1.96(±0.09)              & 2.28(±0.17)              & 4.20(±0.20)               & 6.28(±0.40)               & 1.62(±0.04)              & 1.10(±0.17)              & 1.58(±0.17)              & 2.00(±0.15)              & 2.40(±0.13)              & 4.34(±0.14)               & 6.30(±0.50) 
    \\ \hline
    EMCDR-MF                                      & 1.11($\pm$0.06)          & 0.69($\pm$0.06)          & 1.02($\pm$0.18)          & 1.41($\pm$0.19)          & 2.08($\pm$0.32)          & 2.86($\pm$0.48)           & 4.69($\pm$0.84)           & 1.02($\pm$0.13)          & 0.48($\pm$0.06)          & 1.01($\pm$0.24)          & 1.45($\pm$0.13)          & 1.30($\pm$0.06)          & 3.39($\pm$0.05)           & 6.51($\pm$0.32)           \\
    EMCDR-NGCF                                    & 1.22($\pm$0.08)          & 0.91($\pm$0.04)          & 1.35($\pm$0.12)          & 1.58($\pm$0.19)          & 2.86($\pm$0.64)          & 4.43($\pm$0.32)           & 5.73($\pm$0.64)           & 1.06($\pm$0.09)          & 0.46($\pm$0.25)          & 1.02($\pm$0.08)          & 1.52($\pm$0.09)          & 1.30($\pm$0.16)          & 4.17($\pm$0.32)           & 5.73($\pm$0.28)           \\
    PTUPCDR                                       & 1.57($\pm$0.04)          & 1.19($\pm$0.02)          & 1.45($\pm$0.07)          & 2.74($\pm$0.19)          & 3.39($\pm$0.64)          & 5.73($\pm$0.32)           & 7.29($\pm$0.96)           & 2.03($\pm$0.21)          & 1.55($\pm$0.23)          & 2.21($\pm$0.34)          & 3.15($\pm$0.13)          & 2.86($\pm$0.32)          & 5.47($\pm$0.32)           & 9.90($\pm$0.36)           \\ \hline
    DisenCDR                                      & 2.15($\pm$0.02)          & 1.47($\pm$0.16)          & 2.31($\pm$0.09)          & 3.04($\pm$0.35)          & 3.23($\pm$0.34)          & 6.61($\pm$0.37)           & 10.07($\pm$1.27)          & 3.20($\pm$0.54)          & 2.82($\pm$0.27)          & 3.44($\pm$0.17)          & 4.03($\pm$0.69)          & 4.95($\pm$0.84)          & 7.40($\pm$0.46)           & 10.22($\pm$0.70)          \\
    UniCDR                                        & 2.14($\pm$0.12)          & 2.10($\pm$0.19)          & 2.79($\pm$0.21)          & 3.18($\pm$0.25)          & 4.16($\pm$0.50)          & 6.92($\pm$0.61)           & 8.79($\pm$0.77)           & 2.28($\pm$0.11)          & 2.23($\pm$0.12)          & 2.95($\pm$0.17)          & 3.49($\pm$0.22)          & 4.39($\pm$0.25)          & 7.28($\pm$0.44)           & 9.80($\pm$0.71)           \\
    CDRIB                                        & 3.30($\pm$0.08)     & 3.40($\pm$0.07)          & 4.43($\pm$0.10)          & 5.15($\pm$0.08)           & 6.71($\pm$0.08)          & 10.83($\pm$0.18)          & 14.18($\pm$0.09)    & 
    3.58($\pm$0.07)          & 3.63($\pm$0.08)          & 4.70($\pm$0.09)          & 5.48($\pm$0.13)          & 6.86($\pm$0.12)          & 11.10($\pm$0.18)          & 14.79($\pm$0.42)          \\
    DPMCDR                                       & {\ul 3.89($\pm$0.19)}    & {\ul 4.06($\pm$0.24)}    & {\ul 5.26($\pm$0.28)}    & {\ul 6.00($\pm$0.27)}    & {\ul 7.76($\pm$0.44)}    & {\ul 12.54($\pm$0.60)}    & {\ul 15.94($\pm$0.59)}          & {\ul 3.61($\pm$0.10)}    & {\ul 3.72($\pm$0.13)}    & {\ul 4.82($\pm$0.09)}    & {\ul 5.56($\pm$0.10)}    & {\ul 7.19($\pm$0.27)}    & {\ul 11.59($\pm$0.08)}    & {\ul 15.05($\pm$0.13)}    \\ \hline
    \textbf{HJID*}                                          & \textbf{4.10($\pm$0.06)*} & \textbf{4.48($\pm$0.04)*} & \textbf{5.55($\pm$0.05)*} & \textbf{6.28($\pm$0.05)*} & \textbf{8.76($\pm$0.07)*} & \textbf{13.04($\pm$0.18)*} & \textbf{16.45($\pm$0.15)*} & \textbf{3.81($\pm$0.05)*} & \textbf{4.00($\pm$0.09)*} & \textbf{5.14($\pm$0.05)*} & \textbf{5.83($\pm$0.08)*} & \textbf{7.72($\pm$0.20)*} & \textbf{12.27($\pm$0.12)*} & \textbf{16.52($\pm$0.23)*} \\
    Improved                                      & 5.31\%                & 10.44\%               & 5.63\%                & 4.61\%                & 12.92\%               & 4.00\%                 & 3.15\%                 & 5.66\%                & 7.45\%                & 6.59\%                & 4.89\%                & 7.26\%                & 5.92\%                 & 9.74\%                 \\ \hline
    \end{tabular}
    }

    \centering
    \resizebox{\linewidth}{!}{
    \begin{tabular}{l|ccccccc|ccccccc}
    \hline
    \multicolumn{1}{c|}{\multirow{2}{*}{Methods}} & \multicolumn{7}{c|}{Game}                                                                                                                                        & \multicolumn{7}{c}{Sport}                                                                                                                                        \\ \cline{2-15} 
    \multicolumn{1}{c|}{}                         & MRR                  & NDCG@10              & NDCG@20              & NDCG@30              & HR@10                & HR@20                 & HR@30                 & MRR                  & NDCG@10              & NDCG@20              & NDCG@30              & HR@10                & HR@20                 & HR@30                 \\ \hline
    CMF                                           & 1.45($\pm$0.28)          & 1.60($\pm$0.13)          & 2.12($\pm$0.19)          & 2.56($\pm$0.23)          & 4.13($\pm$0.61)          & 5.88($\pm$0.72)           & 8.29($\pm$0.84)           & 1.52($\pm$0.21)          & 1.17($\pm$0.26)          & 1.43($\pm$0.42)          & 1.76($\pm$0.37)          & 2.49($\pm$0.14)          & 4.21($\pm$0.19)           & 5.43($\pm$0.85)           \\
    LFM                                           & 1.48(±0.05)              & 1.02(±0.04)              & 1.46(±0.16)              & 1.84(±0.12)              & 2.42(±0.15)              & 4.16(±0.25)               & 5.94(±0.33)               & 1.56(±0.11)              & 1.06(±0.14)              & 1.52(±0.24)              & 1.96(±0.28)              & 2.44(±0.24)              & 4.32(±0.27)               & 6.40(±0.54) 
    \\ \hline
    EMCDR-MF                                      & 1.58($\pm$0.35)          & 1.60($\pm$0.13)          & 2.12($\pm$0.19)          & 2.12($\pm$0.54)          & 3.34($\pm$0.15)          & 5.09($\pm$0.19)           & 5.99($\pm$0.23)           & 1.39($\pm$0.25)          & 1.44($\pm$0.21)          & 1.89($\pm$0.47)          & 2.50($\pm$0.52)          & 3.01($\pm$0.32)          & 5.83($\pm$0.28)           & 7.36($\pm$0.71)           \\
    EMCDR-NGCF                                    & 1.35($\pm$0.27)          & 0.97($\pm$0.33)          & 1.62($\pm$0.50)          & 2.12($\pm$0.54)          & 2.68($\pm$0.28)          & 7.95($\pm$0.66)           & 8.63($\pm$0.54)           & 1.86($\pm$0.06)          & 1.15($\pm$0.09)          & 1.66($\pm$0.29)          & 2.11($\pm$0.36)          & 2.86($\pm$0.14)          & 4.95($\pm$0.08)           & 7.36($\pm$0.27)           \\
    PTUPCDR                                       & 1.54($\pm$0.30)          & 1.35($\pm$0.14)          & 2.05($\pm$0.06)          & 2.27($\pm$0.08)          & 2.42($\pm$0.09)          & 5.02($\pm$0.55)           & 6.92($\pm$0.14)           & 1.56($\pm$0.23)          & 1.33($\pm$0.14)          & 2.45($\pm$0.20)          & 2.78($\pm$0.02)          & 2.90($\pm$0.18)          & 7.02($\pm$0.37)           & 8.94($\pm$0.50)           \\ \hline
    DisenCDR                                      & 2.18($\pm$0.08)          & 1.50($\pm$0.03)          & 2.80($\pm$0.38)          & 3.63($\pm$0.57)          & 3.32($\pm$0.19)          & 8.52($\pm$1.59)           & 12.42($\pm$1.43)          & 2.33($\pm$0.21)          & 1.72($\pm$0.22)          & 2.82($\pm$0.39)          & 3.69($\pm$0.48)          & 3.87($\pm$0.60)          & 8.28($\pm$1.28)           & 12.40($\pm$1.72)          \\
    UniCDR                                        & 2.65($\pm$0.11)          & 2.62($\pm$0.08)          & 3.35($\pm$0.06)          & 3.92($\pm$0.10)          & 4.95($\pm$0.03)          & 7.87($\pm$0.46)           & 10.56($\pm$0.51)          & 2.60($\pm$0.42)          & 2.59($\pm$0.50)          & 3.38($\pm$0.58)          & 3.83($\pm$0.60)          & 4.94($\pm$0.85)          & 8.07($\pm$1.16)           & 10.21($\pm$1.24)          \\
    CDRIB                                        & {\ul 3.64($\pm$0.10)}    & {\ul 3.88($\pm$0.10)}    & {\ul 4.97($\pm$0.11)}    & {\ul 5.71($\pm$0.09)}    & {\ul 7.71($\pm$0.08)}    & {\ul 12.04($\pm$0.15)}    & {\ul 15.52($\pm$0.15)}    & {\ul 4.25($\pm$0.29)}    & {\ul 4.50($\pm$0.29)}    & {\ul 5.83($\pm$0.36)}    & {\ul 6.64($\pm$0.44)}    & {\ul 8.67($\pm$0.42)}    & {\ul 13.95($\pm$0.72)}    & {\ul 17.76($\pm$1.07)}    \\
    DPMCDR                                       & 3.43($\pm$0.09)          & 3.61($\pm$0.12)          & 4.82($\pm$0.08)          & 5.58($\pm$0.08)          & 7.28($\pm$0.24)          & 12.07($\pm$0.24)          & 15.67($\pm$0.38)          & 4.01($\pm$0.22)          & 4.22($\pm$0.38)          & 5.56($\pm$0.25)          & 6.39($\pm$0.21)          & 8.36($\pm$0.43)          & 13.69($\pm$0.31)          & 17.63($\pm$0.16)          \\ \hline
    \textbf{HJID*}                                          & \textbf{3.92($\pm$0.06)*} & \textbf{4.17($\pm$0.09)*} & \textbf{5.23($\pm$0.06)*} & \textbf{5.97($\pm$0.09)*} & \textbf{7.96($\pm$0.18)*} & \textbf{12.22($\pm$0.07)*} & \textbf{15.70($\pm$0.26)*} & \textbf{4.60($\pm$0.10)*} & \textbf{4.91($\pm$0.10)*} & \textbf{6.32($\pm$0.08)*} & \textbf{7.20($\pm$0.07)*} & \textbf{9.45($\pm$0.21)*} & \textbf{15.05($\pm$0.08)*} & \textbf{19.22($\pm$0.19)*} \\
    Improved                                      & 7.81\%                & 7.31\%                & 5.33\%                & 4.63\%                & 3.21\%                & 1.52\%                 & 1.12\%                 & 8.04\%                & 8.98\%                & 8.36\%                & 8.52\%                & 8.94\%                & 7.93\%                 & 8.25\%                 \\ \hline
    \end{tabular}
    }
    
    \centering
    \resizebox{\linewidth}{!}{
    \begin{tabular}{l|ccccccc|ccccccc}
    \hline
    \multicolumn{1}{c|}{\multirow{2}{*}{Methods}} & \multicolumn{7}{c|}{Video}                                                                                                                                       & \multicolumn{7}{c}{Cloth}                                                                                                                                        \\ \cline{2-15} 
    \multicolumn{1}{c|}{}                         & MRR                  & NDCG@10              & NDCG@20              & NDCG@30              & HR@10                & HR@20                 & HR@30                 & MRR                  & NDCG@10              & NDCG@20              & NDCG@30              & HR@10                & HR@20                 & HR@30                 \\ \hline
    CMF                                           & 1.01($\pm$0.04)          & 0.45($\pm$0.14)          & 0.70($\pm$0.09)          & 1.20($\pm$0.11)          & 1.04($\pm$0.32)          & 2.08($\pm$0.64)           & 4.43($\pm$0.84)           & 1.20($\pm$0.13)          & 0.73($\pm$0.07)          & 1.45($\pm$0.20)          & 1.89($\pm$0.14)          & 1.82($\pm$0.32)          & 4.69($\pm$0.55)           & 6.77($\pm$0.32)           \\
    LFM                                           & 1.55(±0.08)              & 1.07(±0.08)              & 1.53(±0.17)              & 1.93(±0.13)              & 2.54(±0.21)              & 4.37(±0.31)               & 6.23(±0.30)               & 1.64(±0.12)              & 1.11(±0.13)              & 1.59(±0.15)              & 2.06(±0.25)              & 2.56(±0.16)              & 4.536(±0.27)              & 6.72(±0.46) 
    \\ \hline
    EMCDR-MF                                      & 0.87($\pm$0.49)          & 0.49($\pm$0.14)          & 0.80($\pm$0.09)          & 1.19($\pm$0.49)          & 1.56($\pm$0.32)          & 2.86($\pm$0.55)           & 4.69($\pm$0.32)           & 1.01($\pm$0.09)          & 0.70($\pm$0.07)          & 1.03($\pm$0.21)          & 1.36($\pm$0.17)          & 2.08($\pm$0.32)          & 3.39($\pm$0.84)           & 4.95($\pm$0.64)           \\
    EMCDR-NGCF                                    & 2.11($\pm$0.29)          & 1.63($\pm$0.36)          & 1.88($\pm$0.43)          & 2.27($\pm$0.45)          & 2.34($\pm$0.55)          & 3.39($\pm$0.84)           & 6.25($\pm$0.91)           & 1.00($\pm$0.12)          & 0.51($\pm$0.03)          & 1.15($\pm$0.16)          & 1.53($\pm$0.24)          & 1.56($\pm$0.09)          & 4.17($\pm$0.32)           & 5.99($\pm$0.84)           \\
    PTUPCDR                                       & 1.86($\pm$0.15)          & 1.77($\pm$0.19)          & 2.01($\pm$0.21)          & 2.52($\pm$0.29)          & 3.91($\pm$0.55)          & 4.95($\pm$0.78)           & 7.29($\pm$1.09)           & 1.45($\pm$0.04)          & 1.37($\pm$0.06)          & 2.01($\pm$0.33)          & 2.23($\pm$0.27)          & 3.65($\pm$0.32)          & 6.25($\pm$0.59)           & 7.29($\pm$0.91)           \\ \hline
    DisenCDR                                      & 2.12($\pm$0.04)          & 1.50($\pm$0.10)          & 2.49($\pm$0.07)          & 3.30($\pm$0.07)          & 3.45($\pm$0.26)          & 7.40($\pm$0.15)           & 11.23($\pm$0.26)          & 2.16($\pm$0.01)          & 1.60($\pm$0.05)          & 2.44($\pm$0.04)          & 3.26($\pm$0.05)          & 3.53($\pm$0.12)          & 6.88($\pm$0.16)           & 10.78($\pm$0.31)          \\
    UniCDR                                        & 1.71($\pm$0.19)          & 1.63($\pm$0.21)          & 2.14($\pm$0.20)          & 2.48($\pm$0.21)          & 3.29($\pm$0.32)          & 5.32($\pm$0.26)           & 6.92($\pm$0.31)           & 2.93($\pm$0.16)          & 2.94($\pm$0.18)          & 3.89($\pm$0.25)          & 4.57($\pm$0.27)          & 5.84($\pm$0.39)          & 9.60($\pm$0.67)           & 12.82($\pm$0.77)          \\
    CDRIB                                        & {\ul 4.13($\pm$0.09)}    & {\ul 4.30($\pm$0.07)}    & {\ul 5.70($\pm$0.11)}    & {\ul 6.65($\pm$0.13)}    & {\ul 8.56($\pm$0.07)}    & {\ul 14.18($\pm$0.20)}    & {\ul 18.63($\pm$0..36)}   & {\ul 3.42($\pm$0.11)}    & {\ul 3.52($\pm$0.14)}    & {\ul 4.54($\pm$0.10)}    & {\ul 5.26($\pm$0.11)}    & {\ul 6.84($\pm$0.24)}    & {\ul 10.79($\pm$0.12)}    & {\ul 14.21($\pm$0.06)}    \\
    DPMCDR                                       & 3.94($\pm$0.13)          & 4.07($\pm$0.16)          & 5.44($\pm$0.25)          & 6.45($\pm$0.25)          & 8.27($\pm$0.34)          & 13.76($\pm$0.72)          & 18.48($\pm$0.69)          & 3.31($\pm$0.26)          & 3.32($\pm$0.30)          & 4.40($\pm$0.26)          & 5.16($\pm$0.21)          & 6.41($\pm$0.39)          & 10.59($\pm$0.28)          & 14.17($\pm$0.16)          \\ \hline
    \textbf{HJID*}                                          & \textbf{4.56($\pm$0.08)*} & \textbf{4.82($\pm$0.08)*} & \textbf{6.19($\pm$0.08)*} & \textbf{7.15($\pm$0.07)*} & \textbf{9.46($\pm$0.06)*} & \textbf{14.92($\pm$0.10)*} & \textbf{19.42($\pm$0.14)*} & \textbf{3.69($\pm$0.22)*} & \textbf{3.82($\pm$0.22)*} & \textbf{4.82($\pm$0.24)*} & \textbf{5.55($\pm$0.21)*} & \textbf{7.22($\pm$0.17)*} & \textbf{11.22($\pm$0.29)*} & \textbf{14.66($\pm$0.20)*} \\
    Improved                                      & 10.27\%               & 12.20\%               & 8.59\%                & 7.52\%                & 10.46\%               & 5.24\%                 & 4.25\%                 & 7.85\%                & 8.44\%                & 6.21\%                & 5.46\%                & 5.53\%                & 3.96\%                 & 3.18\%                 \\ \hline
    \end{tabular}
    }
    
    \centering
    \resizebox{\linewidth}{!}{
    \begin{tabular}{l|ccccccc|ccccccc}
    \hline
    \multicolumn{1}{c|}{\multirow{2}{*}{Methods}} & \multicolumn{7}{c|}{Video}                                                                                                                                        & \multicolumn{7}{c}{Sport}                                                                                                                                        \\ \cline{2-15} 
    \multicolumn{1}{c|}{}                         & MRR                  & NDCG@10              & NDCG@20              & NDCG@30              & HR@10                 & HR@20                 & HR@30                 & MRR                  & NDCG@10              & NDCG@20              & NDCG@30              & HR@10                & HR@20                 & HR@30                 \\ \hline
    CMF                                           & 1.32($\pm$0.41)          & 0.85($\pm$0.22)          & 1.12($\pm$0.17)          & 1.45($\pm$0.15)          & 1.56($\pm$0.28)           & 2.60($\pm$0.24)           & 4.17($\pm$0.28)           & 0.97($\pm$0.09)          & 0.45($\pm$0.04)          & 0.85($\pm$0.04)          & 1.01($\pm$0.02)          & 1.04($\pm$0.05)          & 2.60($\pm$0.17)           & 3.39($\pm$0.22)           \\ 
    LFM                                           & 1.31(±0.07)              & 0.90(±0.10)              & 1.29(±0.19)              & 1.63(±0.21)              & 2.14(±0.18)               & 3.68(±0.25)               & 5.26(±0.26)               & 1.38(±0.20)              & 0.94(±0.24)              & 1.33(±0.15)              & 1.73(±0.15)              & 2.15(±0.15)              & 3.82(±0.27)               & 5.66(±0.28) 
    \\ \hline
    EMCDR-MF                                      & 1.12($\pm$0.22)          & 0.53($\pm$0.21)          & 0.73($\pm$0.13)          & 1.02($\pm$0.08)          & 1.04($\pm$0.05)           & 1.82($\pm$0.06)           & 3.13($\pm$0.08)           & 0.99($\pm$0.07)          & 0.66($\pm$0.02)          & 0.97($\pm$0.13)          & 1.25($\pm$0.23)          & 1.82($\pm$0.32)          & 3.13($\pm$0.32)           & 5.47($\pm$0.46)           \\
    EMCDR-NGCF                                    & 1.16($\pm$0.13)          & 0.77($\pm$0.14)          & 0.96($\pm$0.14)          & 1.41($\pm$0.21)          & 1.82($\pm$0.32)           & 2.60($\pm$0.32)           & 4.69($\pm$0.55)           & 1.07($\pm$0.22)          & 0.62($\pm$0.31)          & 1.14($\pm$0.37)          & 1.47($\pm$0.41)          & 1.56($\pm$0.55)          & 3.65($\pm$0.59)           & 5.21($\pm$0.61)           \\
    PTUPCDR                                       & 1.43($\pm$0.37)          & 1.11($\pm$0.32)          & 1.55($\pm$0.35)          & 1.82($\pm$0.36)          & 2.60($\pm$0.32)           & 4.43($\pm$0.64)           & 5.73($\pm$0.55)           & 1.42($\pm$0.20)          & 0.82($\pm$0.25)          & 2.37($\pm$0.22)          & 2.59($\pm$0.11)          & 2.16($\pm$0.32)          & 3.46($\pm$0.55)           & 6.77($\pm$0.69)           \\ \hline
    DisenCDR                                      & 2.22($\pm$0.09)          & 1.34($\pm$0.28)          & 2.31($\pm$0.29)          & 3.27($\pm$0.19)          & 2.98($\pm$0.62)           & 6.88($\pm$0.63)           & 11.42($\pm$0.27)          & 2.35($\pm$0.34)          & 1.99($\pm$0.49)          & 3.15($\pm$0.59)          & 3.86($\pm$0.44)          & 4.66($\pm$1.02)          & 9.25($\pm$1.42)           & 12.61($\pm$0.76)          \\
    UniCDR                                        & 2.97($\pm$0.11)          & 3.06($\pm$0.10)          & 3.81($\pm$0.21)          & 4.42($\pm$0.30)          & 5.94($\pm$0.26)           & 8.89($\pm$0.73)           & 11.74($\pm$1.17)          & 2.95($\pm$0.20)          & 2.97($\pm$0.23)          & 3.89($\pm$0.28)          & 4.52($\pm$0.28)          & 5.80($\pm$0.40)          & 9.46($\pm$0.62)           & 12.45($\pm$0.68)          \\
    CDRIB                                        & {\ul 4.74($\pm$0.19)}    & {\ul 5.04($\pm$0.24)}    & {\ul 6.44($\pm$0.32)}    & {\ul 7.40($\pm$0.37)}    & {\ul 9.73($\pm$0.53)}     & {\ul 15.26($\pm$0.89)}    & 19.79($\pm$1.10)          & {\ul 4.56($\pm$0.12)}    & {\ul 4.88($\pm$0.13)}    & {\ul 6.08($\pm$0.15)}    & {\ul 6.92($\pm$0.21)}    & {\ul 9.17($\pm$0.23)}    & {\ul 13.91($\pm$0.34)}    & {\ul 17.89($\pm$0.64)}    \\
    DPMCDR                                       & 4.57($\pm$0.25)          & 4.81($\pm$0.29)          & 6.28($\pm$0.28)          & 7.28($\pm$0.27)          & 9.42($\pm$0.38)           & 15.25($\pm$0.35)          & {\ul 19.94($\pm$0.33)}    & 4.39($\pm$0.08)          & 4.62($\pm$0.15)          & 5.87($\pm$0.15)          & 6.74($\pm$0.12)          & 8.64($\pm$0.32)          & 13.67($\pm$0.37)          & 17.78($\pm$0.23)          \\ \hline
    \textbf{HJID*}                                         & \textbf{5.05($\pm$0.27)*} & \textbf{5.53($\pm$0.33)*} & \textbf{6.95($\pm$0.29)*} & \textbf{7.96($\pm$0.29)*} & \textbf{10.86($\pm$0.49)*} & \textbf{16.54($\pm$0.32)*} & \textbf{21.33($\pm$0.34)*} & \textbf{4.91($\pm$0.10)*} & \textbf{5.28($\pm$0.09)*} & \textbf{6.59($\pm$0.11)*} & \textbf{7.46($\pm$0.08)*} & \textbf{9.90($\pm$0.05)*} & \textbf{15.13($\pm$0.19)*} & \textbf{19.22($\pm$0.06)*} \\
    Improved                                      & 6.56\%                & 9.55\%                & 7.96\%                & 7.67\%                & 11.64\%                & 8.39\%                 & 6.99\%                 & 7.56\%                & 8.11\%                & 8.48\%                & 7.81\%                & 7.99\%                & 8.76\%                 & 7.46\%                 \\ \hline
    \end{tabular}
    }
    \vspace{-5mm}
\end{table*}

\vspace{-2mm}
\subsection{Overall performance (RQ1)}
As shown in Table. \ref{tab: overall}, we report the average results (with std) of all CDR tasks over six independent runs.
The best results are \textbf{bolded}, and the second-runnerup results are {\ul underlined}.
Generally speaking, HJID consistently outperforms the state-of-the-art~(SOTA) on all CDR scenarios with significant improvements.
We discuss our model against each of \textbf{Alignment}, \textbf{Bridege}, and \textbf{Constraint} methods below.

\vspace{-2mm}
\subsubsection{Comparison with Alignment methods}
CMF directly shares representations of overlapping users in both domains without considering shared and unique information within each domain.
Unfortunately, this leads to unreliable results in both domains. 
In contrast, LFM models the user's different representations with a latent variable, allowing for certain user-specificity across domains.
While LFM outperforms CMF in most evaluation metrics, these Alignment methods falter owing to the inability to separate domain-shared from domain-specific features.
Meanwhile, they face challenges in achieving reasonable prediction results in both domains simultaneously.
In contrast, HJID builds the model upon the feature hierarchy principle and disentanglement of domain-shared and domain-specific latent factors, enabling improved robustness when encountering distribution shifts across different domains.

\vspace{-2mm}
\subsubsection{Comparison with Bridge methods}
Generally, these models incorporate bespoke architectures to endow them with expressiveness for capturing user behavior.
For instance, EMCDR-NGCF and PTUPCDR employ a graph structure and a meta-network, respectively.
In addition, their knowledge transfer between domains is enabled by constructing a transformation bridge.
Still, these methods indiscriminately transfer the complete user representations, yet fail to account for how domain-shared and specific information should be separated.
According to the empirical results, we observe that the performance of PTUPCDR fluctuates across different runs, hampering consistent model efficacy.
Similarly, HJID demonstrates the significance of identifying transferable facet from non-transferable factors in comparison with these Bridge methods.

\subsubsection{Comparison with Constraint methods}
Empirical evidence decisively favors Constraint methods over Alignment and Bridge counterparts, highlighting the utility of optimization constraints in aligning domain-shared attributes.
As an example, DPMCDR, the best performer among the Constraint family, significantly leads PTUPCDR by 4.53\% and 4.24\% in terms of NDCG@30 for the Game-Video CDR task.
However, these methods have yet to demonstrate consistent robustness across all CDR tasks, especially when domains are weakly correlated, such as Game-Sport.
In particular, DisenCDR experiences performance variances of 1.43\% and 1.72\% with HR@30.
To address this, CDRIB and DPMCDR employ VIB to constrain the user/item representations that effectively concentrate on domain-shared information.
Nevertheless, none of these methods have paid sufficient attention to the remaining attributes, i.e., the domain-specific factors grounded in user representations.
In contrast, HJID seeks to address both stable, domain-shared and variant, domain-specific factors with a causally-informed disentanglement.
As a result, HJID demonstrates the best performances among all CDR tasks, regardless of their correlations.

\begin{figure}
    \centering
    \begin{subfigure}[b]{.44\linewidth}
        \centering
        \includegraphics[width=\textwidth]{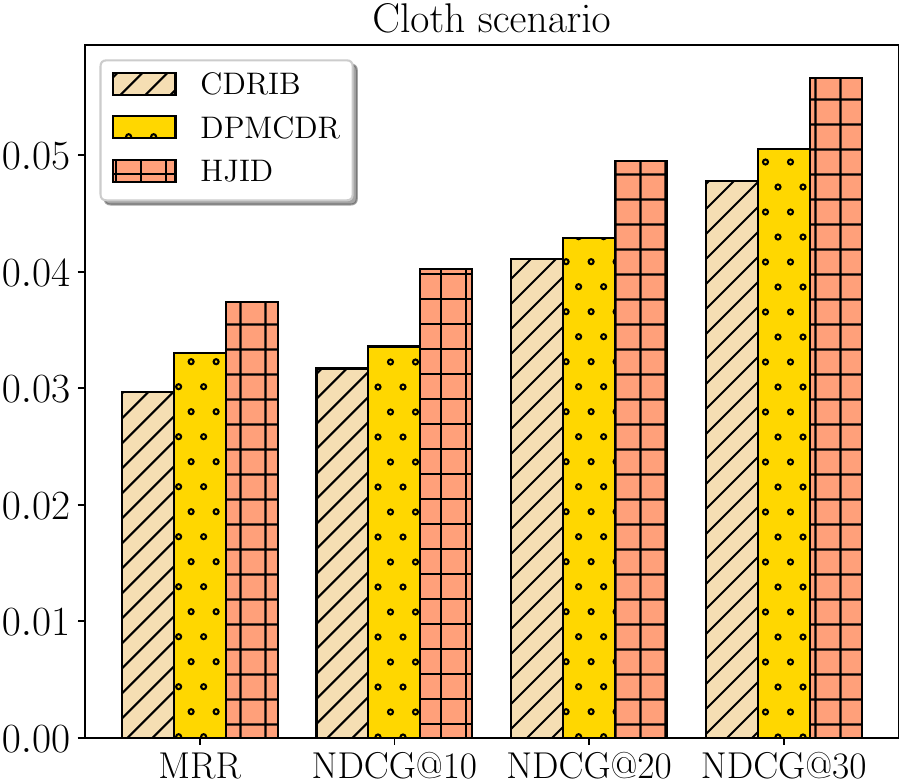}
        \label{fig:non_cs_c}
    \end{subfigure}
    \begin{subfigure}[b]{.44\linewidth}
        \centering
        \includegraphics[width=\textwidth]{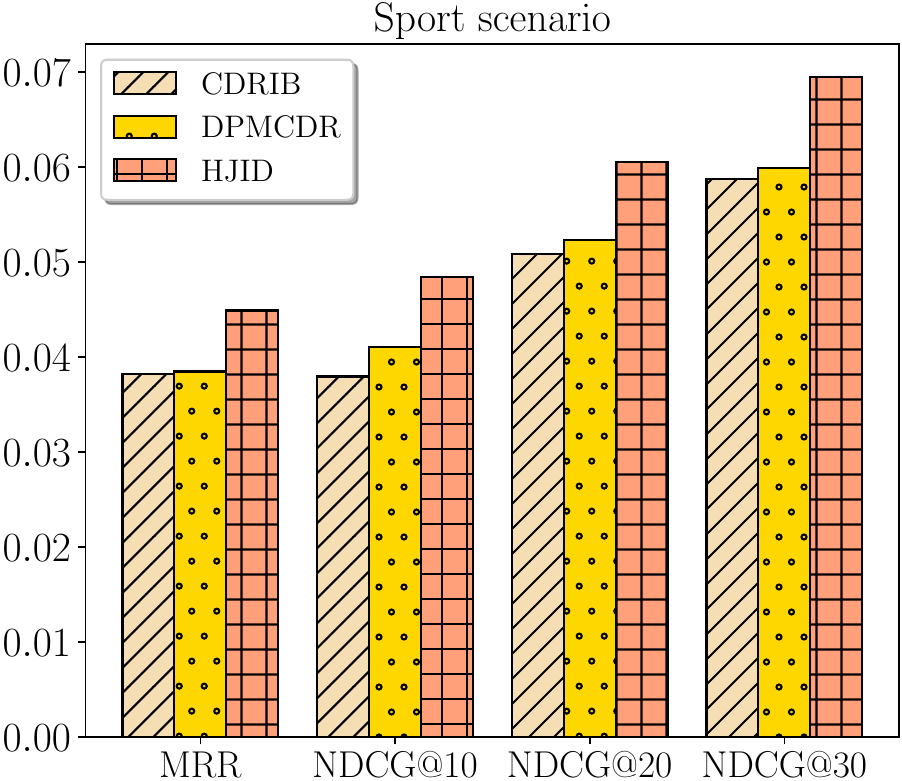}
        \label{fig:non_cs_s}
    \end{subfigure}
    
    \begin{subfigure}[b]{.44\linewidth}
        \centering
        \includegraphics[width=\textwidth]{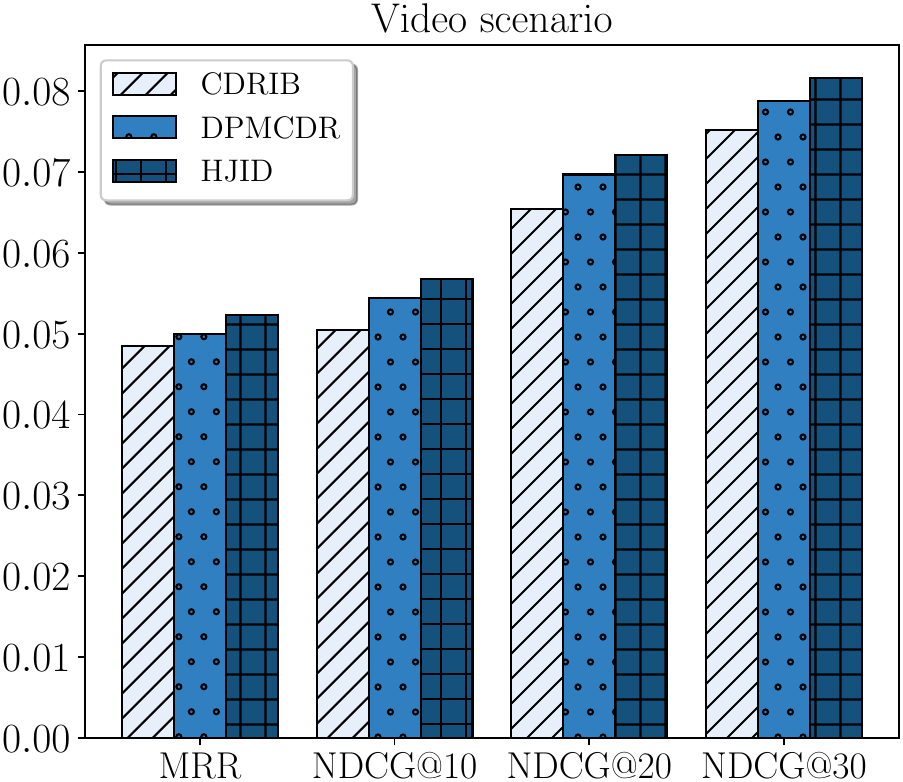}
        \label{fig:non_vs_v}
    \end{subfigure}
    \begin{subfigure}[b]{.44\linewidth}
        \centering
        \includegraphics[width=\textwidth]{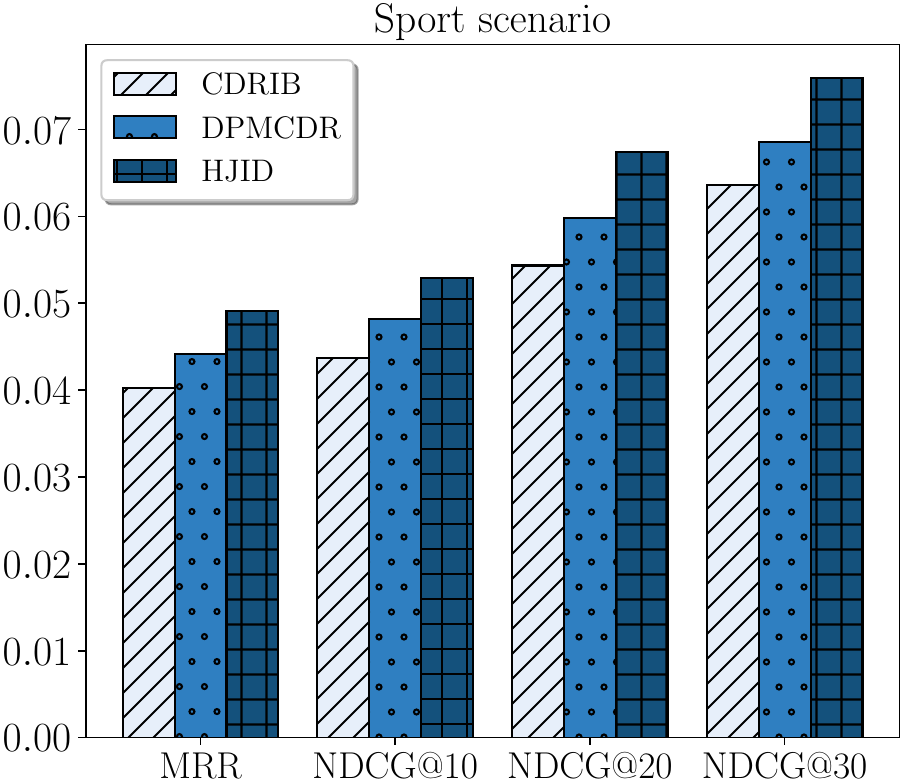}
        \label{fig:non_vs_s}
    \end{subfigure}
    
    \begin{subfigure}[b]{.44\linewidth}
        \centering
        \includegraphics[width=\textwidth]{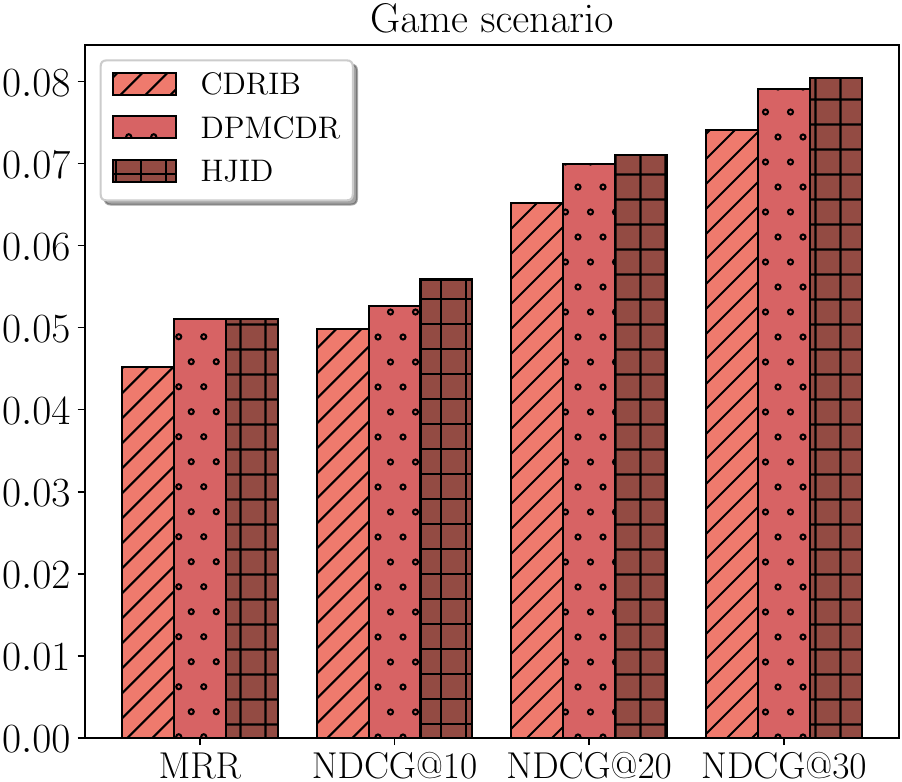}
    \end{subfigure}
    \begin{subfigure}[b]{.44\linewidth}
        \centering
        \includegraphics[width=\textwidth]{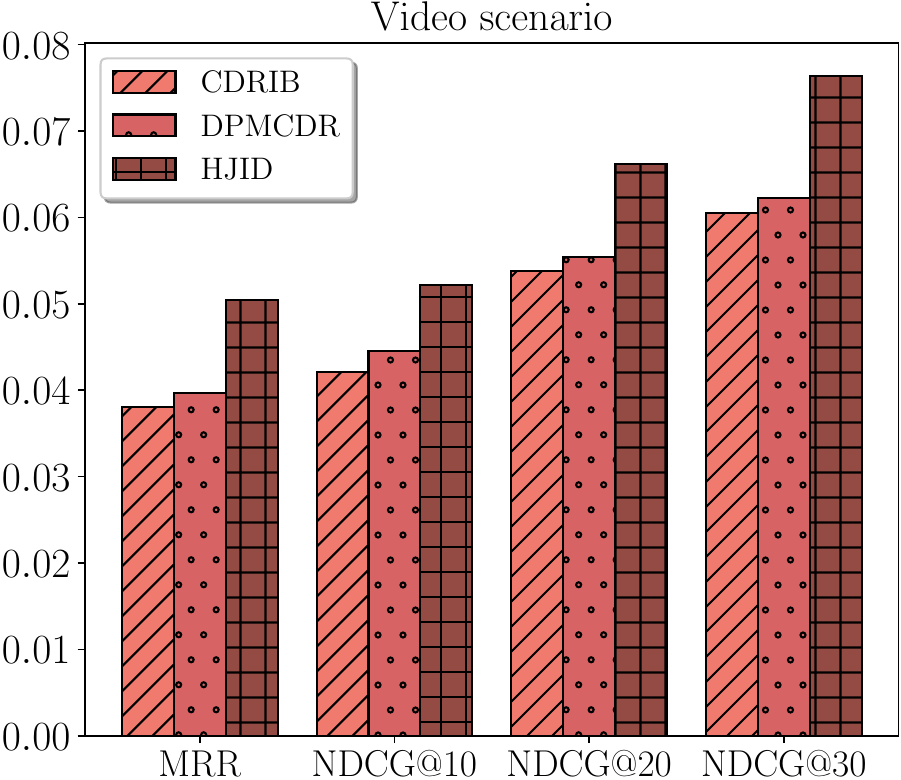}
    \end{subfigure}
    \vspace{-2mm}
    \caption{Non-overlapped CDR comparison.}
    \label{fig: non_overlap}
    \vspace{-5mm}
\end{figure}
\subsection{Non-overlapped CDR scenario (RQ2)}
Non-overlap CDR is an extremely challenging setting, where the model has to successfully extract the common yet adaptable user pattern from observations and make predictions when the training data have no overlapped users between domains.
To answer RQ2, we modify the training dataset to contain only non-overlapped users, and keep the same set of overlapped users, as in the overlapped CDR scenario, for testing and validation. 
Due to the page limitation, we compare HJID with CDRIB and DPMCDR, the top-2 performers among all baselines, for three CDR settings.

Fig.~\ref{fig: non_overlap} illustrates that DPMCDR outperforms CDRIB in all metrics.
Similar to HJID, DPMCDR dispenses with the necessity for exact user overlap as anchors.
Conversely, CDRIB still relies on overlapped users for building cross-domain correspondence.
As the optimization operates over the distributional level, HJID also sidesteps the pairwise correspondence of overlap users, opting instead to enforce causal {\it joint identifiability} to capture cross-domain correlations.
Thanks to the latent disentanglement and {\it joint identifiability}, HJID is able to model both behavioral invariance and variance under domain shifts.

\subsection{Ablation Studies (RQ3)}
To address RQ3, we systematically evaluate the impact of several model variants by removing partial components from the full HJID.
Variant A removes $\boldsymbol{G}$ and performs representation encoding only;
Variant B aligns representations in both shallow and deep subspace simultaneously, without identifying non-transferable factors;
Variant C removes the invertible Flow-based transformations $\boldsymbol{G}(\cdot)$;
Variant D abandons the feature hierarchy principle and applies representation disentanglement for the whole encoder neural network.
Here we report the results with Game-Video, Cloth-Sport, and Game-Cloth in Table.~\ref{tab:ablation} due to page limitation.

\begin{table}
    \caption{Ablation study}
    \label{tab:ablation}
    \resizebox{\linewidth}{!}{
    \begin{tabular}{l|ccc|ccc}
    \hline
    \multirow{2}{*}{Methods} & \multicolumn{3}{c|}{Game}                           & \multicolumn{3}{c}{Video}                           \\ \cline{2-7} 
                             & MRR             & NDCG@30         & HR@30           & MRR             & NDCG@30         & HR@30           \\ \hline
    A                        & 0.0478          & 0.0720          & 0.1840          & 0.0426          & 0.0648          & 0.1704          \\
    B                        & 0.0357          & 0.0620          & 0.1783          & 0.0387          & 0.0604          & 0.1640          \\
    C                        & 0.0488          & 0.0753          & 0.2011          & 0.0393          & 0.0635          & 0.1803          \\ 
    D                        & 0.0452          & 0.0683          & 0.1755          & 0.0418          & 0.0677          & 0.1856          \\ \hline
    HJID                 & \textbf{0.0474} & \textbf{0.0766} & \textbf{0.2026} & \textbf{0.0467} & \textbf{0.0730} & \textbf{0.1958}  \\ \hline
    \end{tabular}
    }

    \resizebox{\linewidth}{!}{
    \begin{tabular}{l|ccc|ccc}
    \hline
    \multirow{2}{*}{Methods} & \multicolumn{3}{c}{Game}                                                 & \multicolumn{3}{c}{Cloth}                           \\ \cline{2-7} 
                             & {MRR}             & NDCG@30         & HR@30           & MRR             & NDCG@30         & HR@30           \\ \hline
    A                        &0.0376          & 0.0586          & 0.1545          & 0.0369          & 0.0556          & 0.1460          \\
    B                        &0.0362          & 0.0566          & 0.1526          & 0.0350          & 0.0533          & 0.1407          \\
    C                        &0.0363          & 0.0576          & 0.1561          & 0.0359          & 0.0536          & 0.1401          \\
    D                        & 0.0381         & 0.0591          & 0.1590          & 0.0355          & 0.0547          & 0.1472          \\ \hline
    HJID                 &\textbf{0.0420} & \textbf{0.0635} & \textbf{0.1645} & \textbf{0.0376} & \textbf{0.0579} & \textbf{0.1651} \\ \hline
    \end{tabular}
    }
    \resizebox{\linewidth}{!}{
    \begin{tabular}{l|ccc|ccc}
    \hline
    \multirow{2}{*}{Methods} & \multicolumn{3}{c}{Cloth}                                     & \multicolumn{3}{c}{Sport}                                 \\ \cline{2-7} 
                             & MRR             & NDCG@30         & HR@30           & MRR             & NDCG@30         & HR@30           \\ \hline
    A                        & 0.0351          & 0.0553          & 0.1504          & 0.0408          & 0.0644          & 0.1755          \\
    B                        & 0.0289          & 0.0485          & 0.1442          & 0.0275          & 0.0404          & 0.1045          \\
    C                        & 0.0281          & 0.0452          & 0.1305          & 0.0441          & 0.0671          & 0.1720          \\
    D                        & 0.0344          & 0.0521          & 0.1372          & 0.0420          & 0.0675          & 0.1849          \\ \hline
    HJID                     & \textbf{0.0383} & \textbf{0.0577} & \textbf{0.1583} & \textbf{0.0458} & \textbf{0.0708} & \textbf{0.1902} \\ \hline
    \end{tabular}
    }
    \vspace{-5mm}
\end{table}

Concretely, variant A can be seen as a baseline without cross-domain adaptation.
Variant B can be interpreted as a member of the Alignment family.
Variant C solely applies MMD to the shallow layers, and thus can be seen as a typical domain adaptation strategy resembling~\cite{long2015learning}.
As shown in Table. \ref{tab:ablation}, in all three tasks, Variant A outperforms Variant B in terms of all metrics.
This may suggest that Alignment methods may be unstable in certain cases, since they inappropriately align non-transferable attributes.
In combating the non-transferability issue, Variant C surpasses Variant B in several key metrics. 
Specifically, its MRR observes a boost of 1.33\% and 0.31\% in the strongly correlated Game-Video task but is barely improved, or even drops in the weakly correlated Game-Cloth and cloth-sport tasks, possibly due to scarce shared information and the absence of an identifiable joint user distribution.
While Variant D ensures {\it joint identifiability}, it may overlook that domain-shared information, resulting in varying performance in some cases, such as Game-Video.
To sum up, the full HJID consistently outperforms all other variants by establishing causal relationships in user representations and ensuring {\it identifiable} joint distribution of çross-user representation, regardless of domain correlation strength.

\vspace{-3mm}
\subsection{Parameter sensitivity (RQ 4)}
\begin{figure*}
    \centering
    \begin{subfigure}[b]{.16\linewidth}
        \centering
        \includegraphics[width=\textwidth]{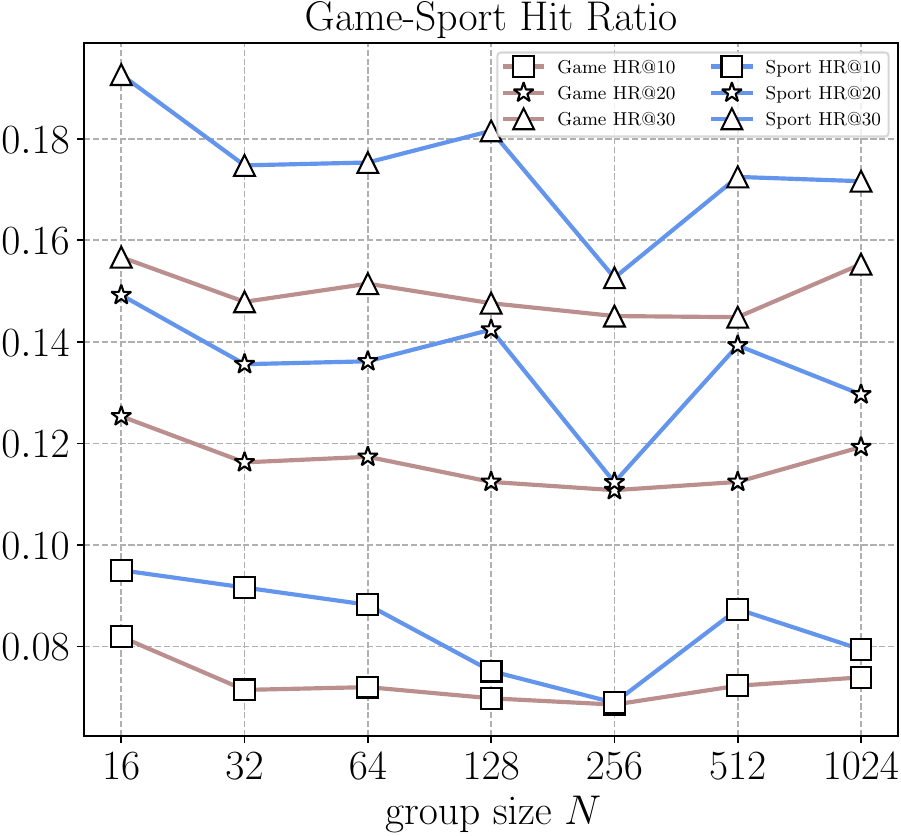}
        \label{fig:gs_groupsize}
    \end{subfigure}
    \begin{subfigure}[b]{.16\linewidth}
        \centering
        \includegraphics[width=\textwidth]{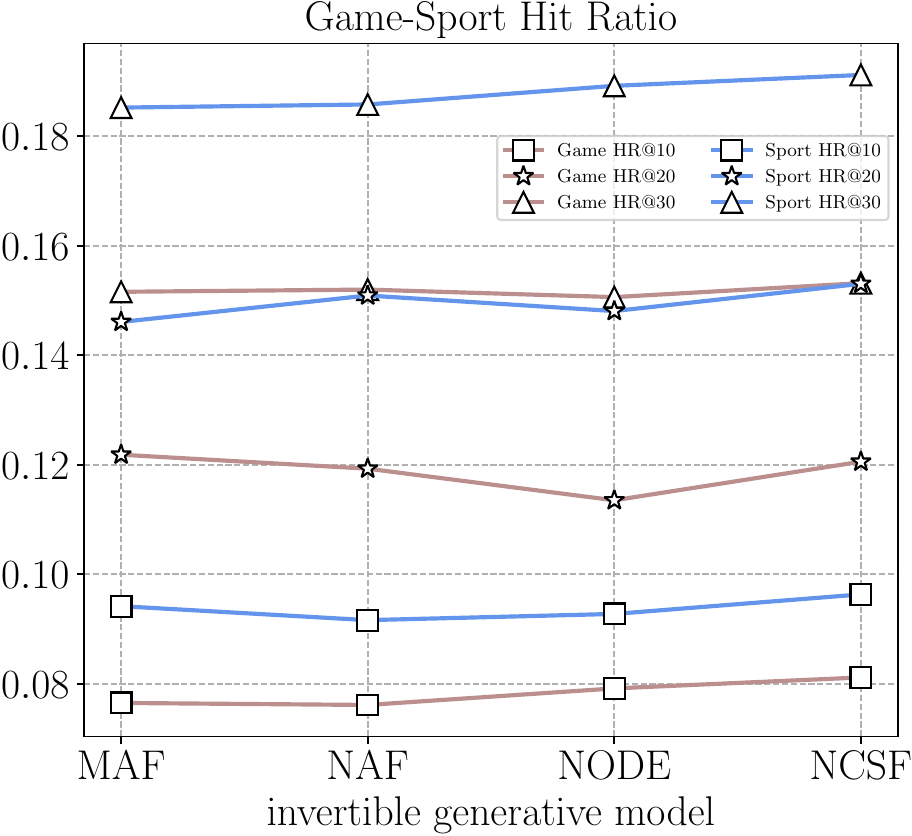}
        \label{fig:gs_flowtype}
    \end{subfigure}
    \begin{subfigure}[b]{.16\linewidth}
        \centering
        \includegraphics[width=\textwidth]{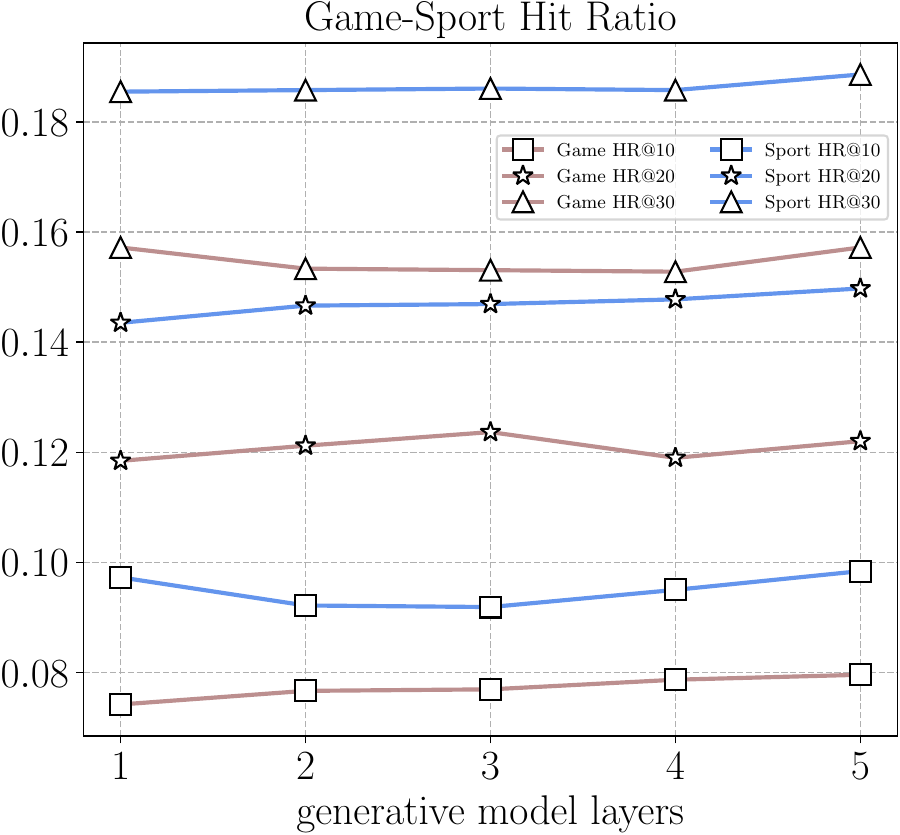}
        \label{fig:gs_flowlayer}
    \end{subfigure}
    \begin{subfigure}[b]{.16\linewidth}
        \centering
        \includegraphics[width=\textwidth]{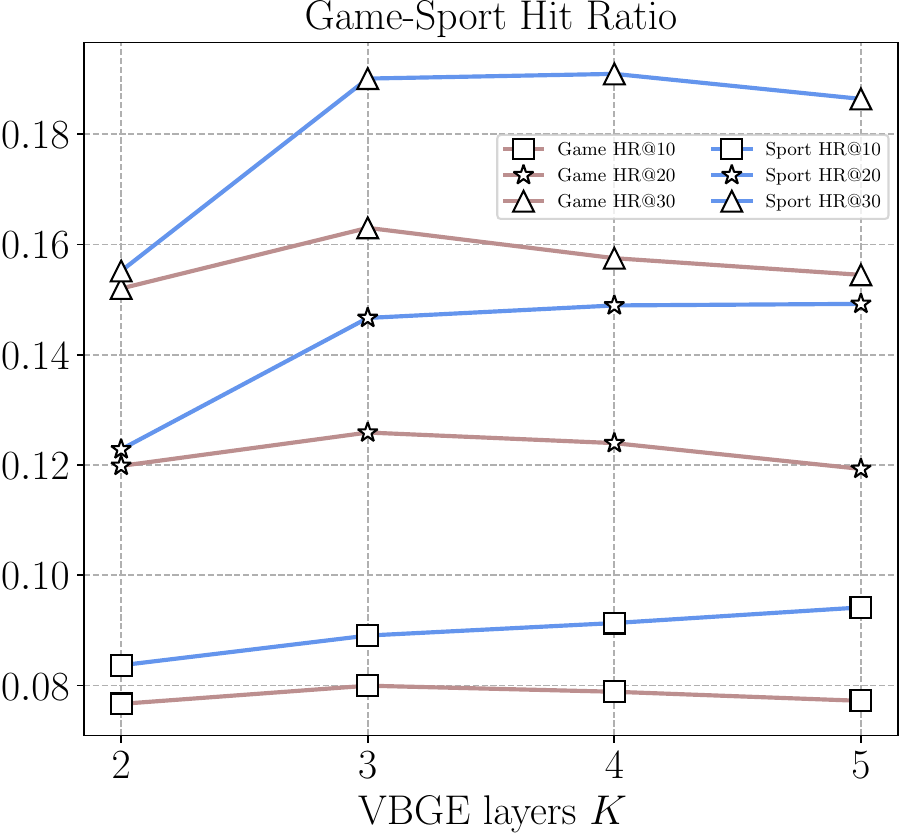}
        \label{fig:gs_align}
    \end{subfigure}
    \begin{subfigure}[b]{.16\linewidth}
        \centering
        \includegraphics[width=\textwidth]{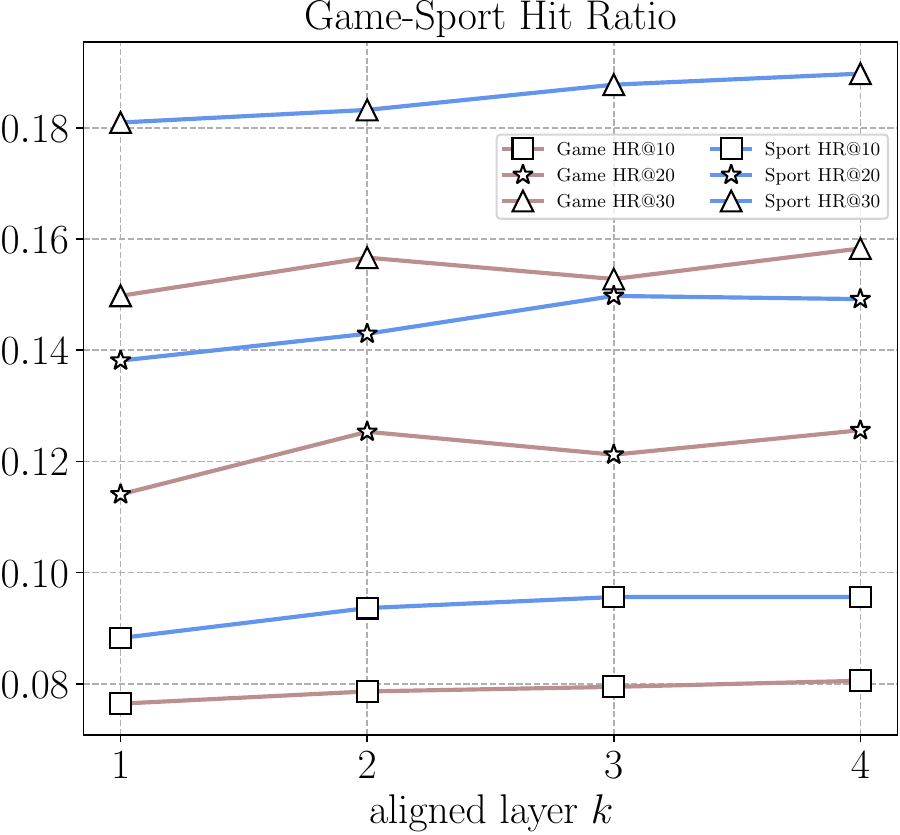}
        \label{fig:gs_gnnlayer}
    \end{subfigure}
    \begin{subfigure}[b]{.16\linewidth}
        \centering
        \includegraphics[width=\textwidth]{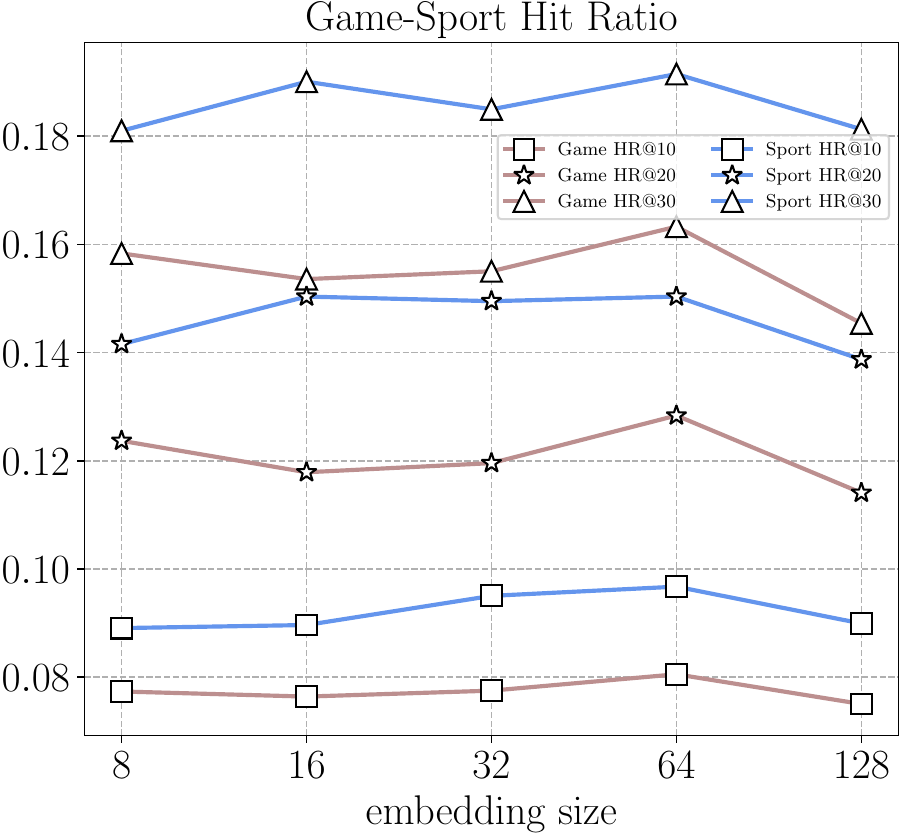}
        \label{fig:gs_embed}
    \end{subfigure}

    \begin{subfigure}[b]{.16\linewidth}
        \centering
        \includegraphics[width=\textwidth]{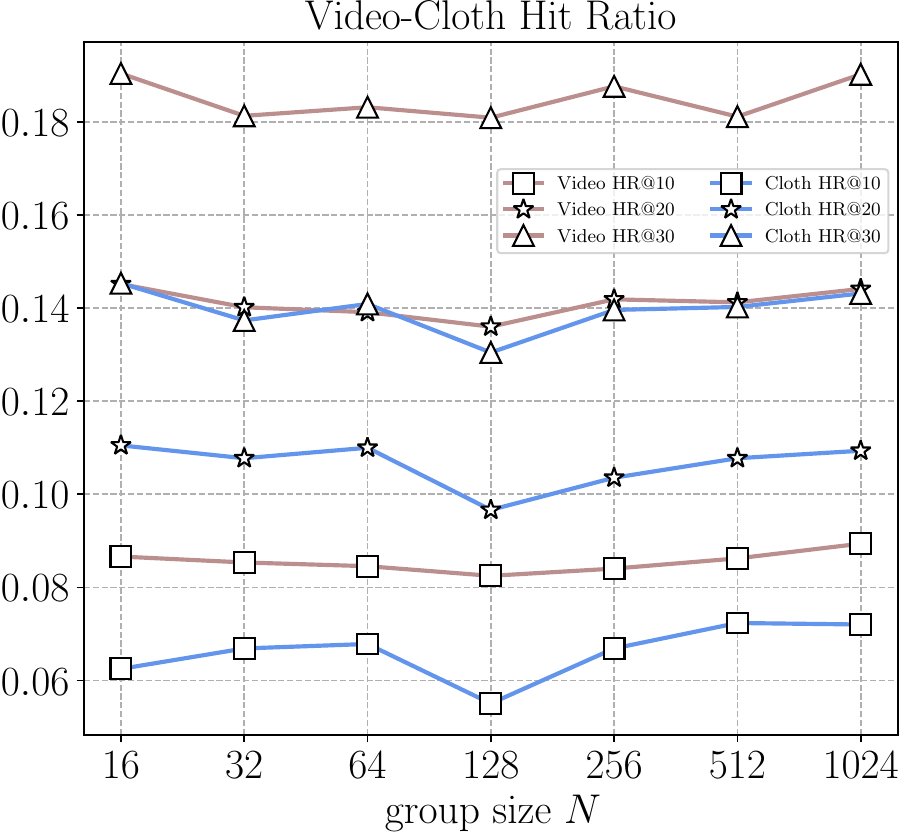}
        \label{fig:vc_groupsize}
    \end{subfigure}
    \begin{subfigure}[b]{.16\linewidth}
        \centering
        \includegraphics[width=\textwidth]{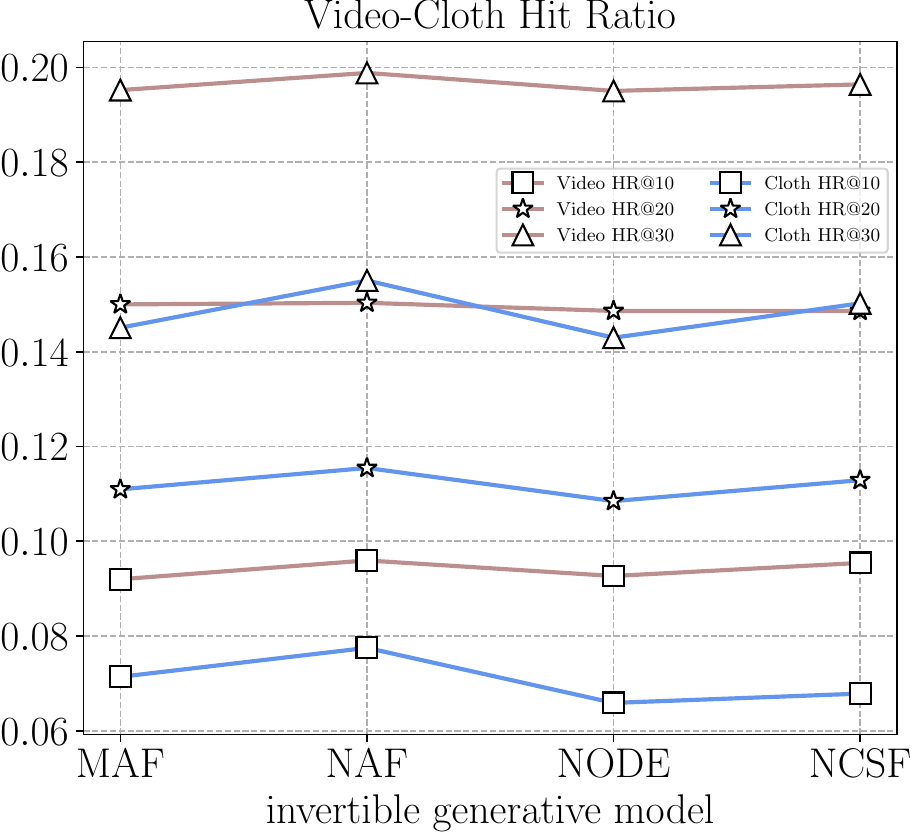}
        \label{fig:vc_flowtype}
    \end{subfigure}
    \begin{subfigure}[b]{.16\linewidth}
        \centering
        \includegraphics[width=\textwidth]{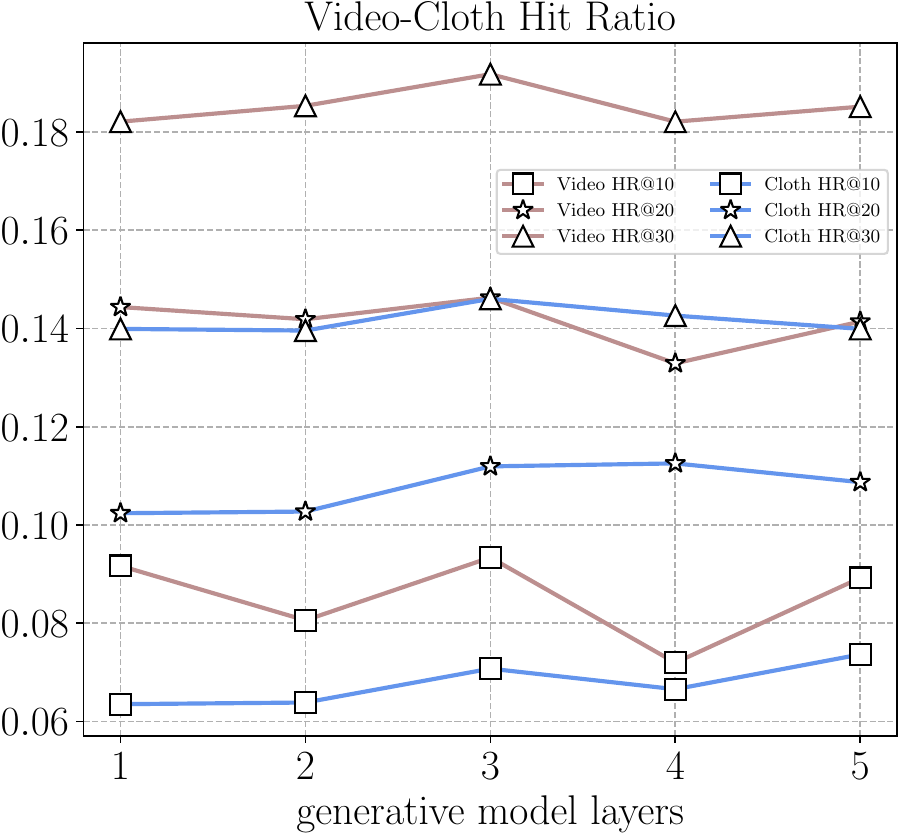}
        \label{fig:vc_flowlayer}
    \end{subfigure}
    \begin{subfigure}[b]{.16\linewidth}
        \centering
        \includegraphics[width=\textwidth]{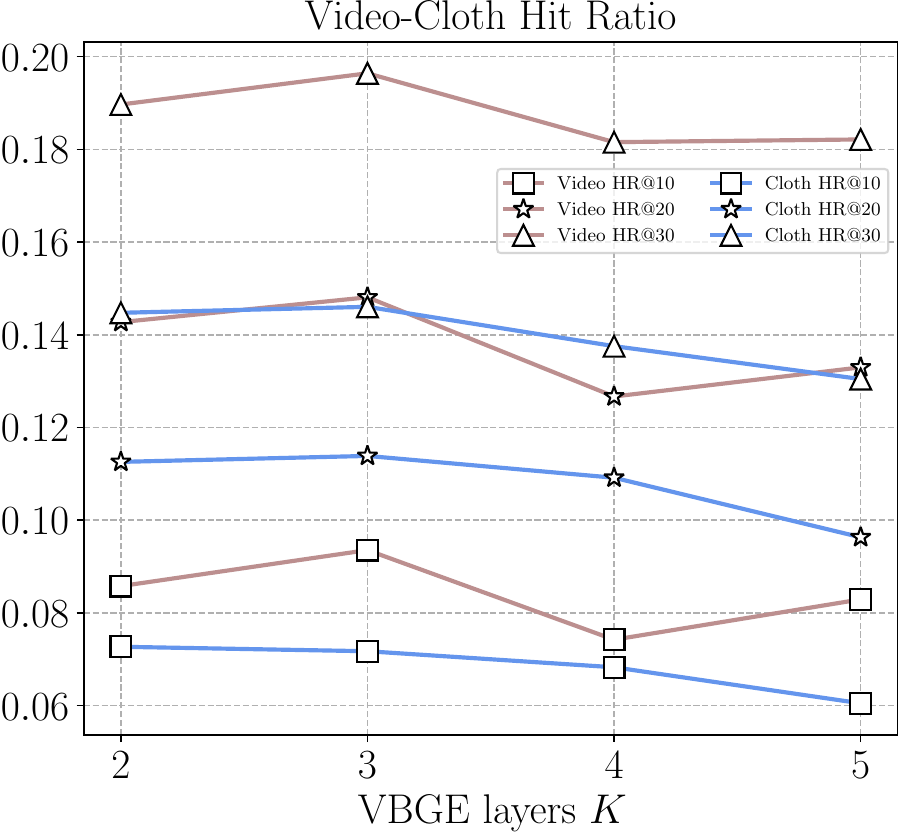}
        \label{fig:vc_align}
    \end{subfigure}
    \begin{subfigure}[b]{.16\linewidth}
        \centering
        \includegraphics[width=\textwidth]{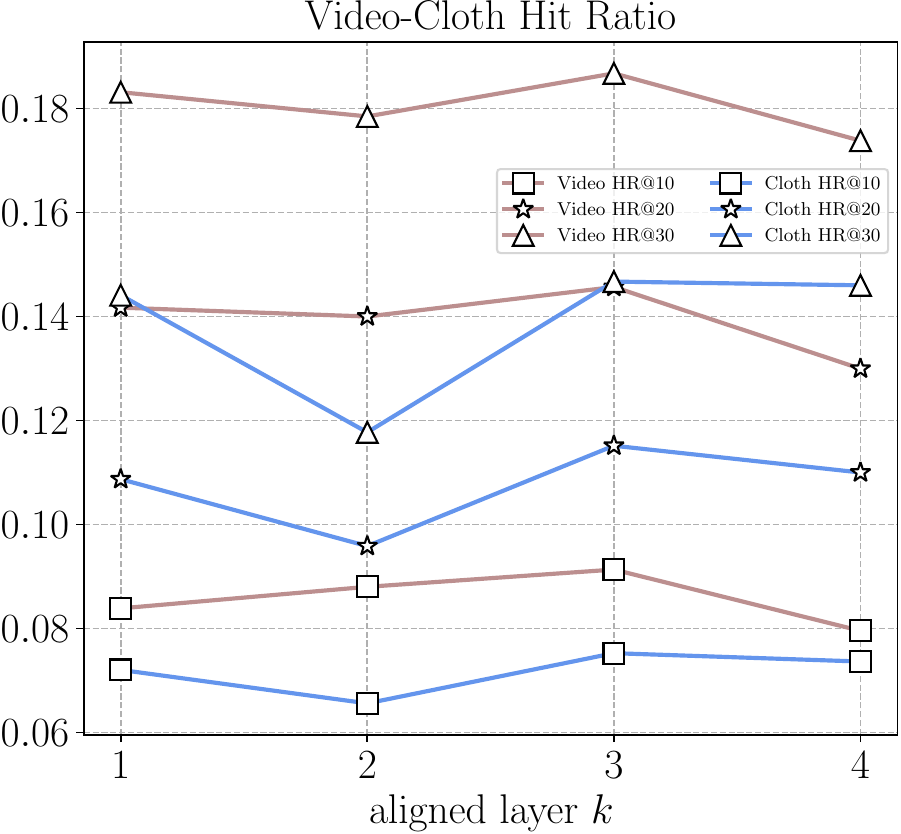}
        \label{fig:vc_gnnlayer}
    \end{subfigure}
    \begin{subfigure}[b]{.16\linewidth}
        \centering
        \includegraphics[width=\textwidth]{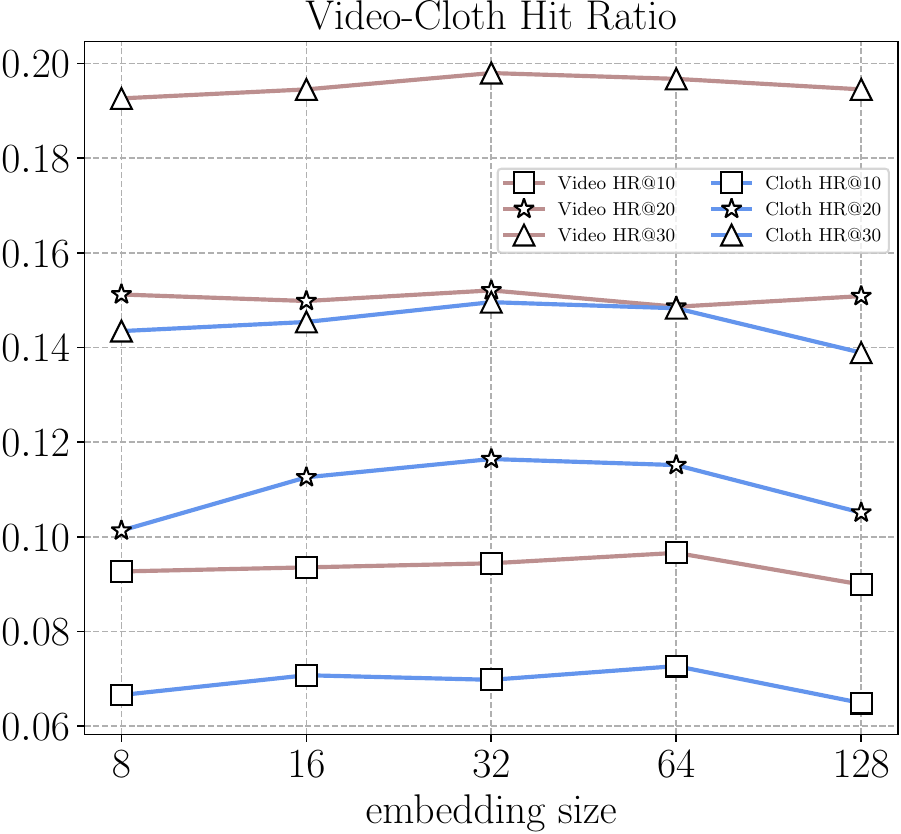}
        \label{fig:vc_embed}
    \end{subfigure}

    \begin{subfigure}[b]{.16\linewidth}
        \centering
        \includegraphics[width=\textwidth]{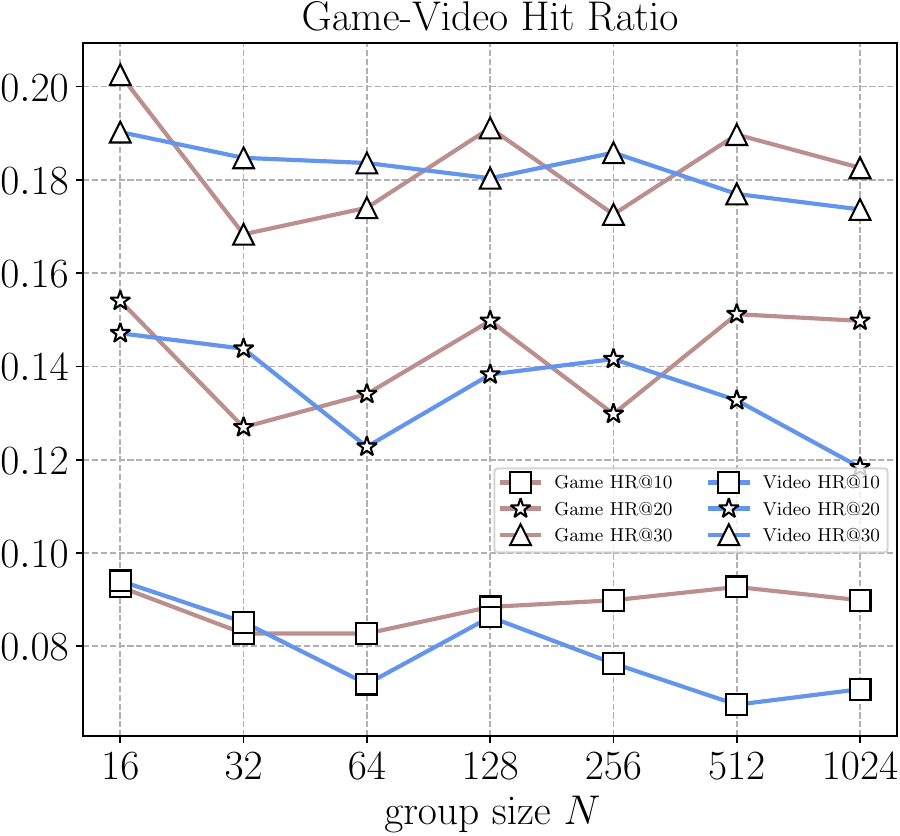}
        \label{fig:gv_groupsize}
    \end{subfigure}
    \begin{subfigure}[b]{.16\linewidth}
        \centering
        \includegraphics[width=\textwidth]{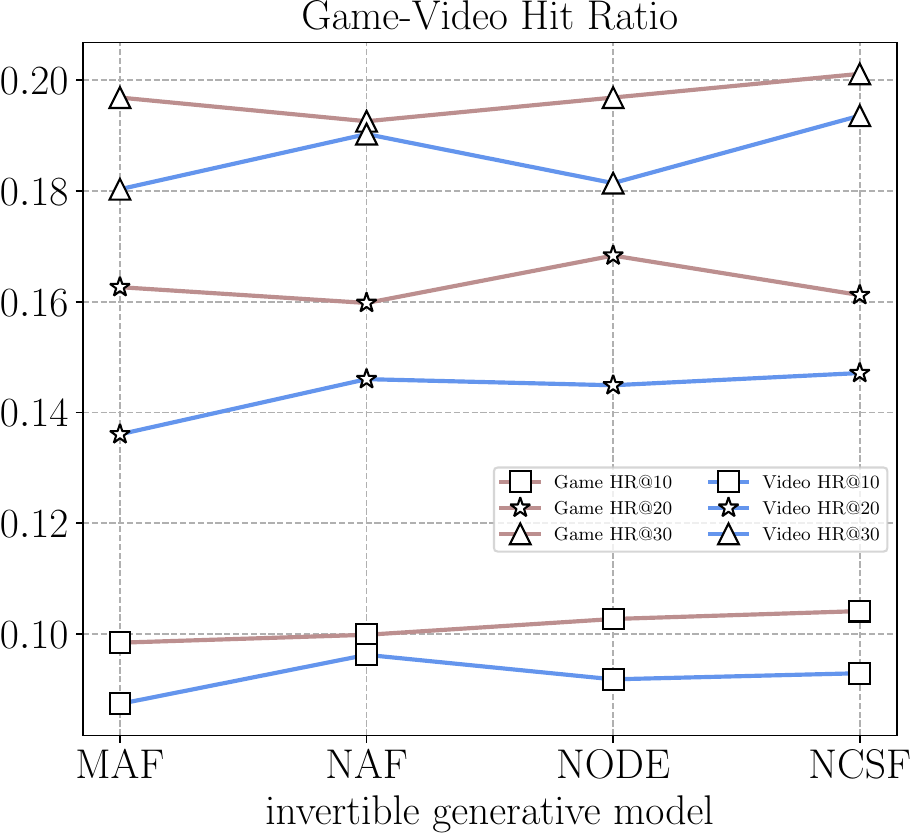}
        \label{fig:gv_flowtype}
    \end{subfigure}
    \begin{subfigure}[b]{.16\linewidth}
        \centering
        \includegraphics[width=\textwidth]{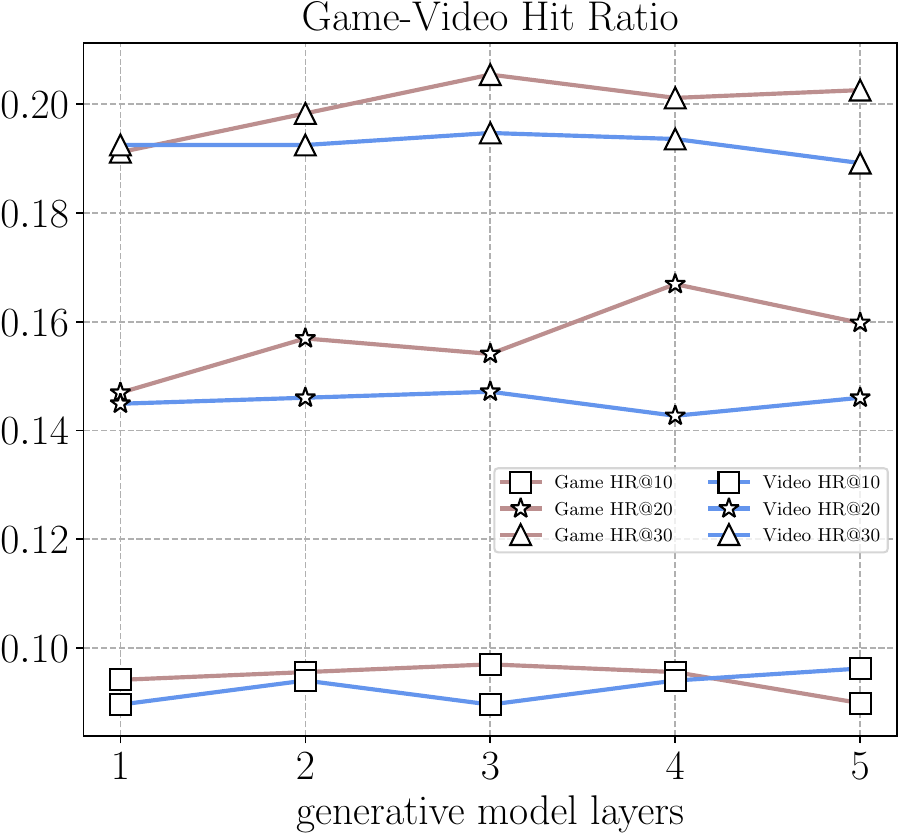}
        \label{fig:gv_flowlayer}
    \end{subfigure}
    \begin{subfigure}[b]{.16\linewidth}
        \centering
        \includegraphics[width=\textwidth]{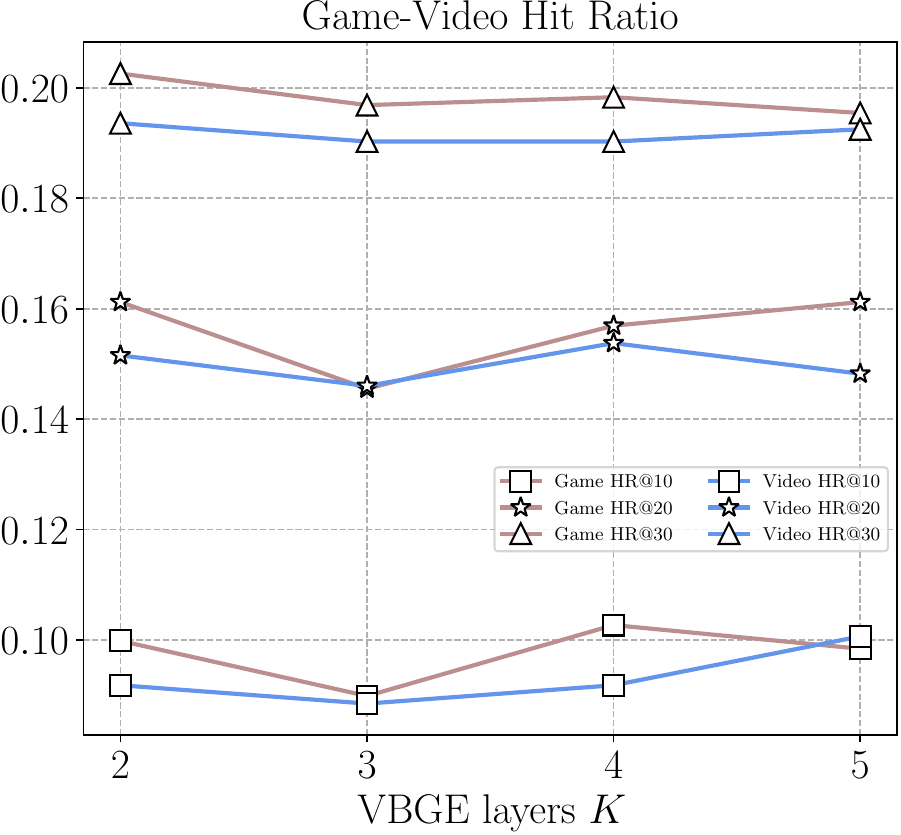}
        \label{fig:gv_align}
    \end{subfigure}
    \begin{subfigure}[b]{.16\linewidth}
        \centering
        \includegraphics[width=\textwidth]{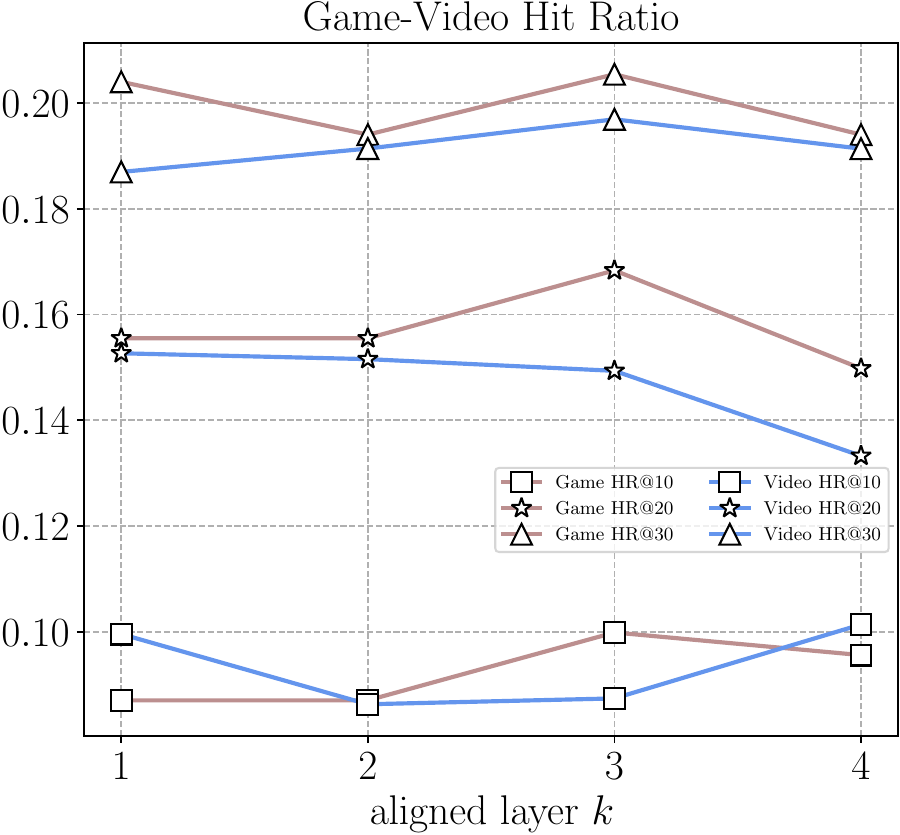}
        \label{fig:gv_gnnlayer}
    \end{subfigure}
    \begin{subfigure}[b]{.16\linewidth}
        \centering
        \includegraphics[width=\textwidth]{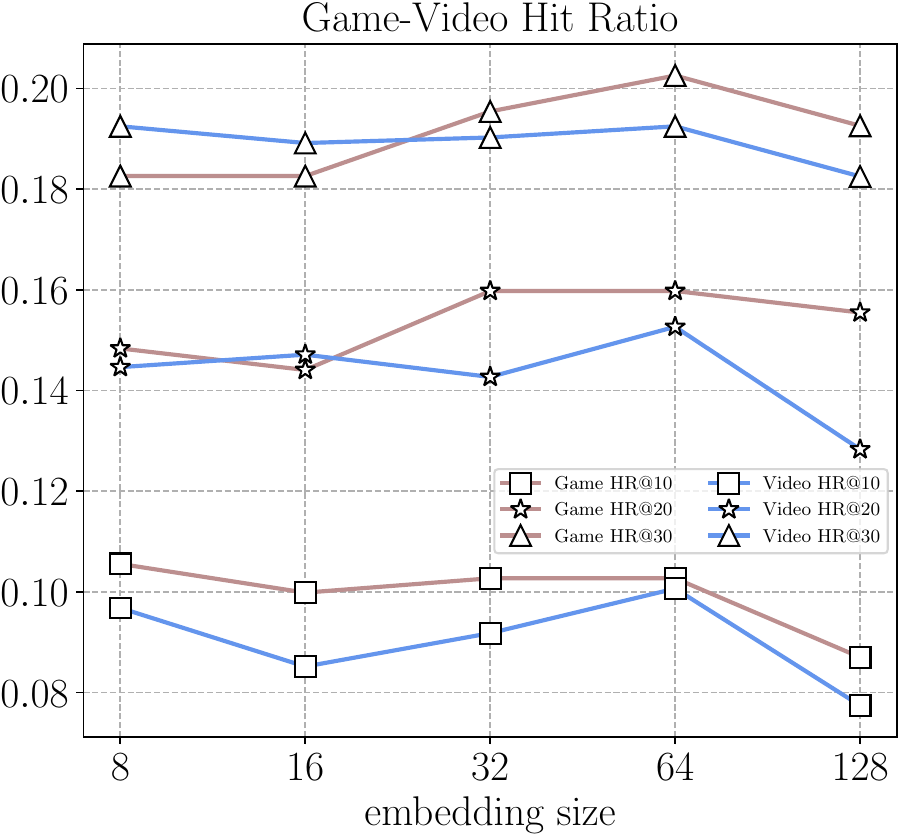}
        \label{fig:gv_embed}
    \end{subfigure}

    \begin{subfigure}[b]{.16\linewidth}
        \centering
        \includegraphics[width=\textwidth]{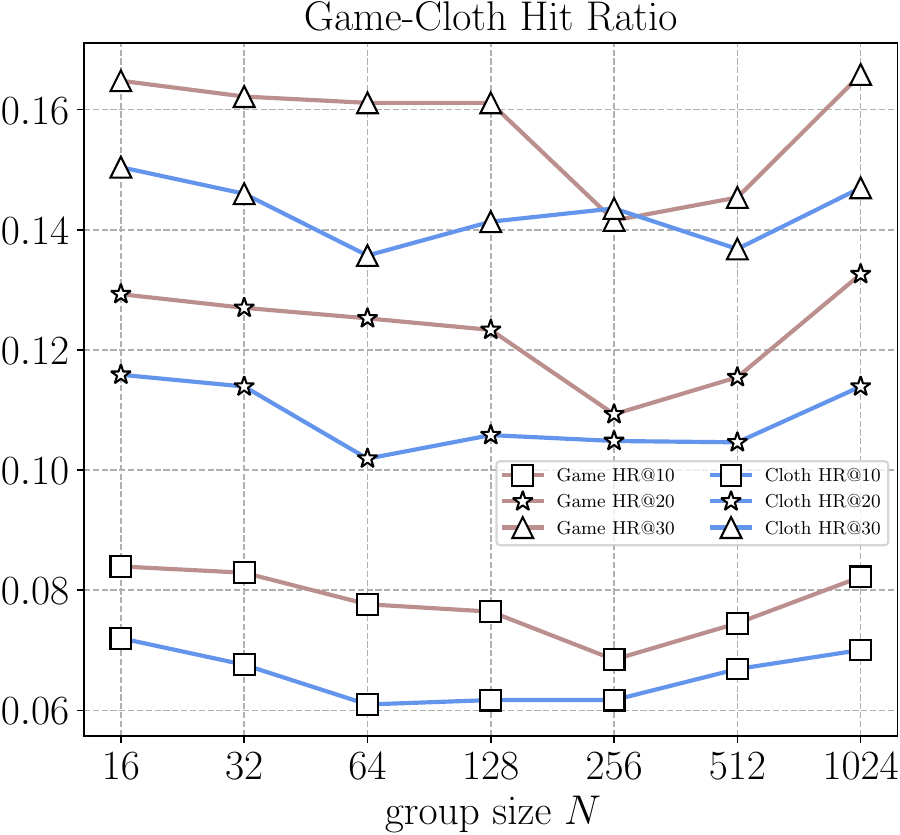}
        \label{fig:gc_groupsize}
    \end{subfigure}
    \begin{subfigure}[b]{.16\linewidth}
        \centering
        \includegraphics[width=\textwidth]{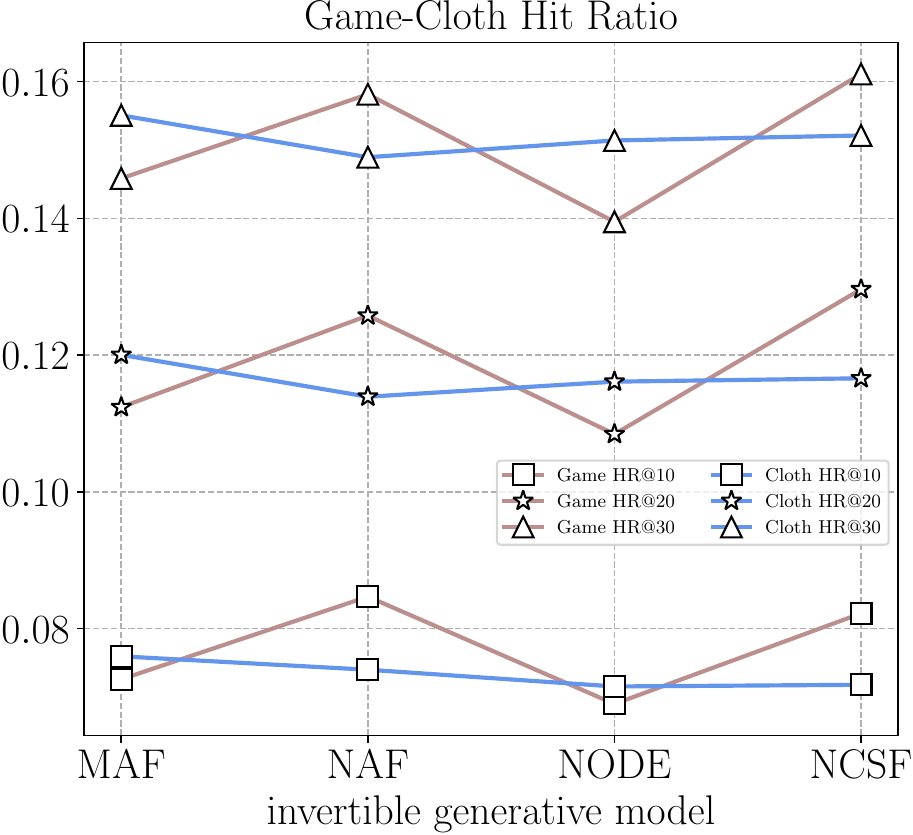}
        \label{fig:gc_flowtype}
    \end{subfigure}
    \begin{subfigure}[b]{.16\linewidth}
        \centering
        \includegraphics[width=\textwidth]{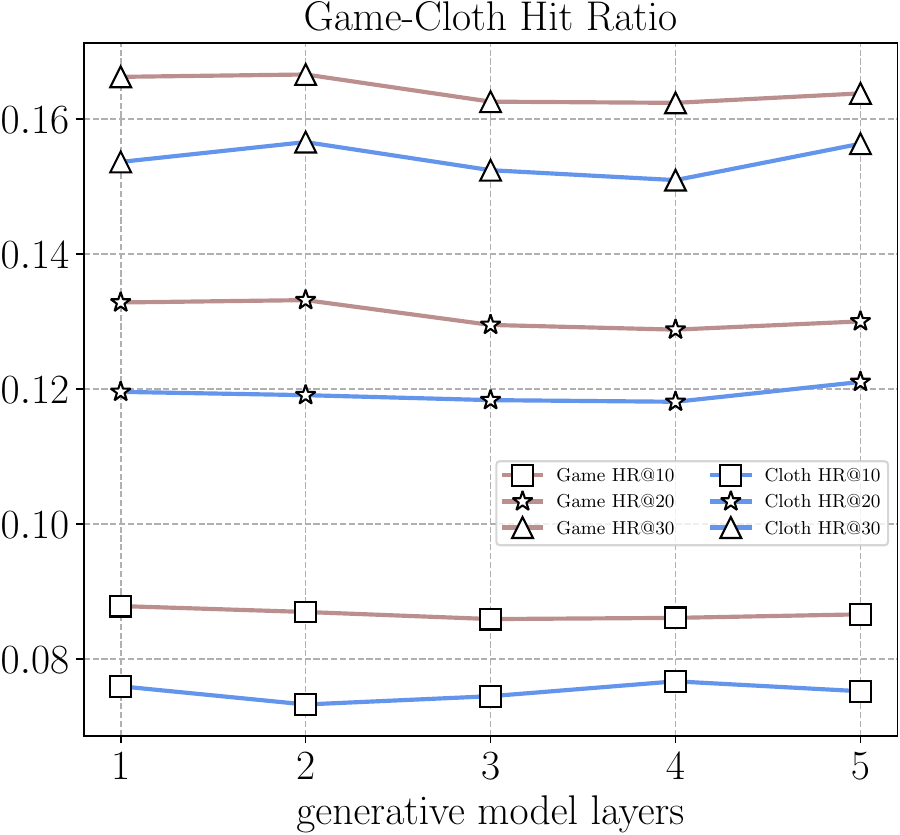}
        \label{fig:gc_flowlayer}
    \end{subfigure}
    \begin{subfigure}[b]{.16\linewidth}
        \centering
        \includegraphics[width=\textwidth]{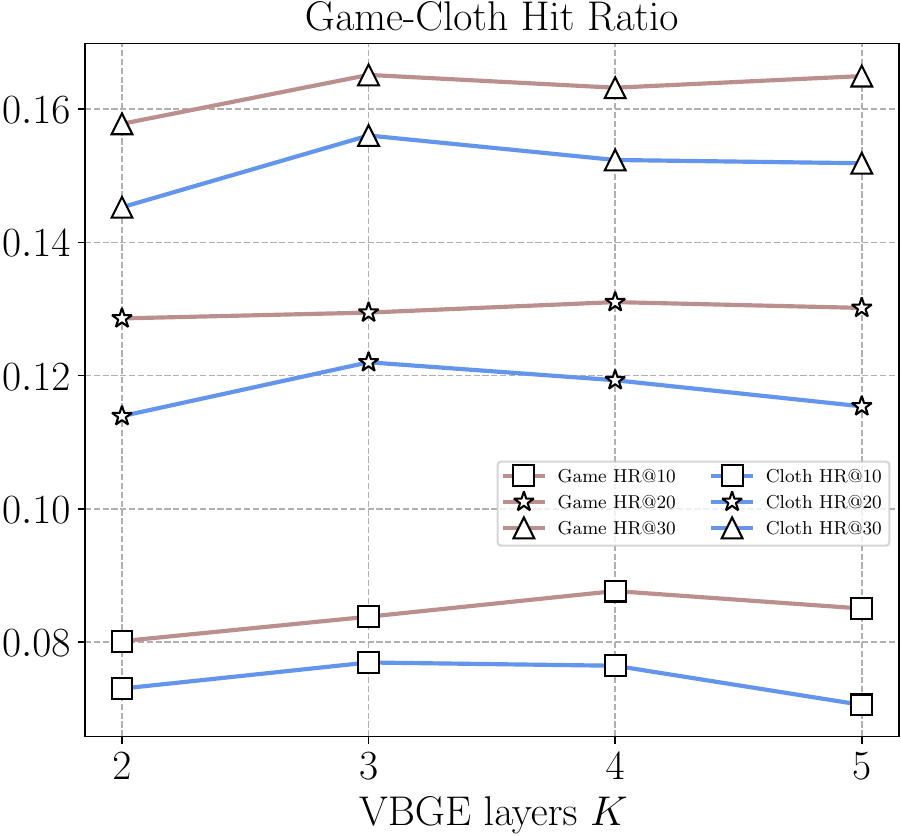}
        \label{fig:gc_align}
    \end{subfigure}
    \begin{subfigure}[b]{.16\linewidth}
        \centering
        \includegraphics[width=\textwidth]{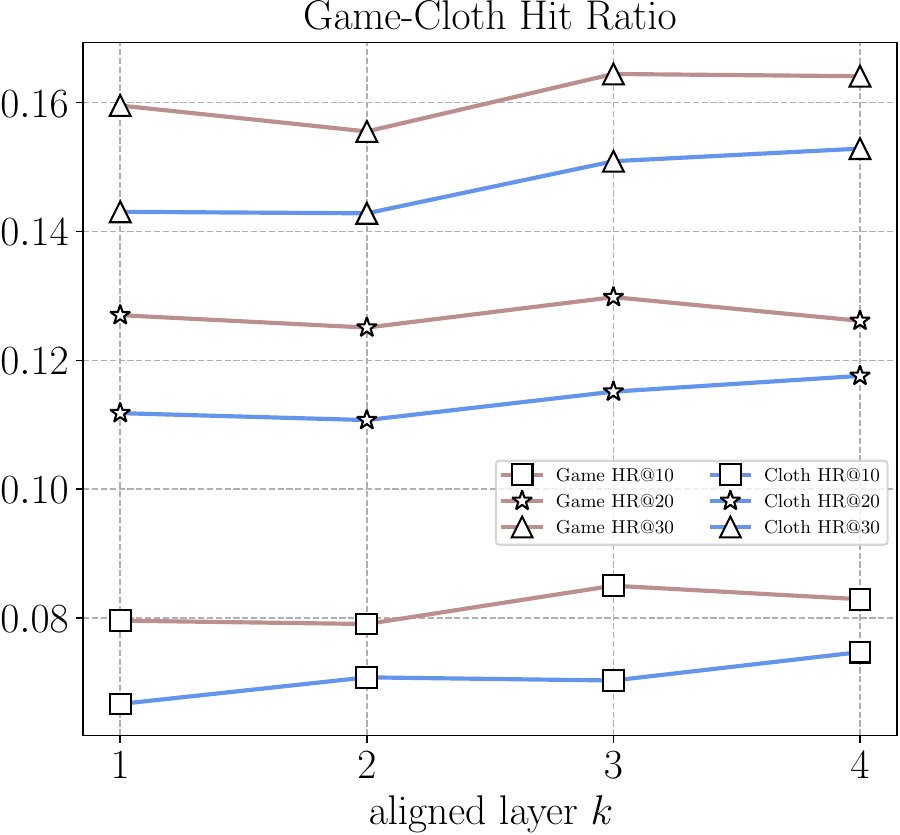}
        \label{fig:gc_gnnlayer}
    \end{subfigure}
    \begin{subfigure}[b]{.16\linewidth}
        \centering
        \includegraphics[width=\textwidth]{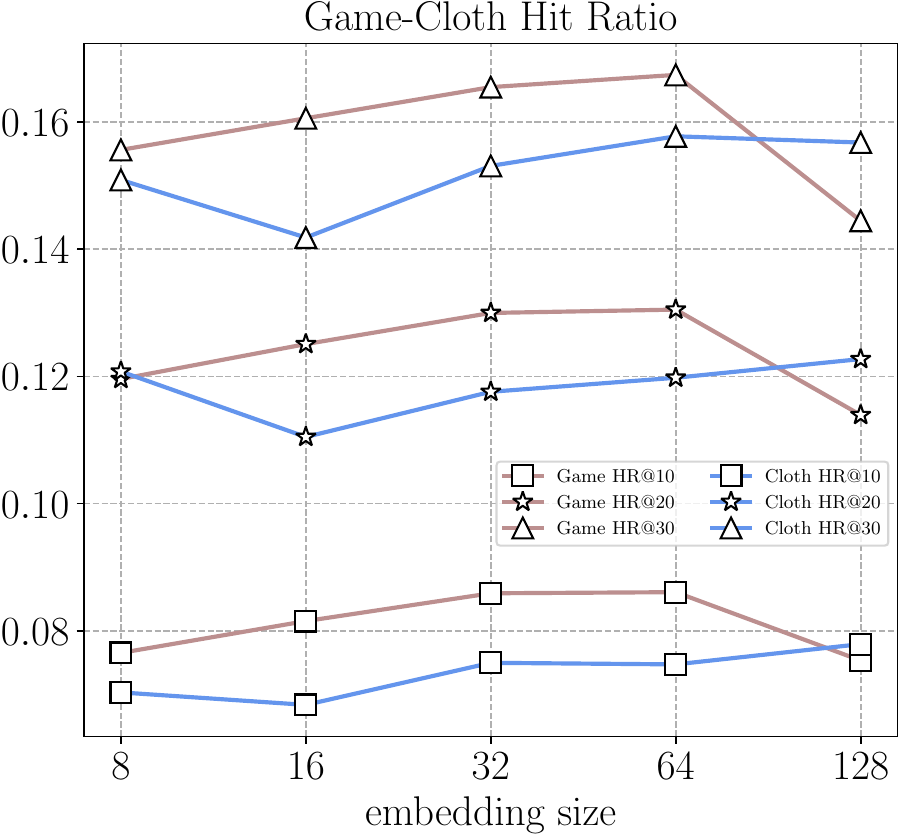}
        \label{fig:gc_embed}
    \end{subfigure}

    \begin{subfigure}[b]{.16\linewidth}
        \centering
        \includegraphics[width=\textwidth]{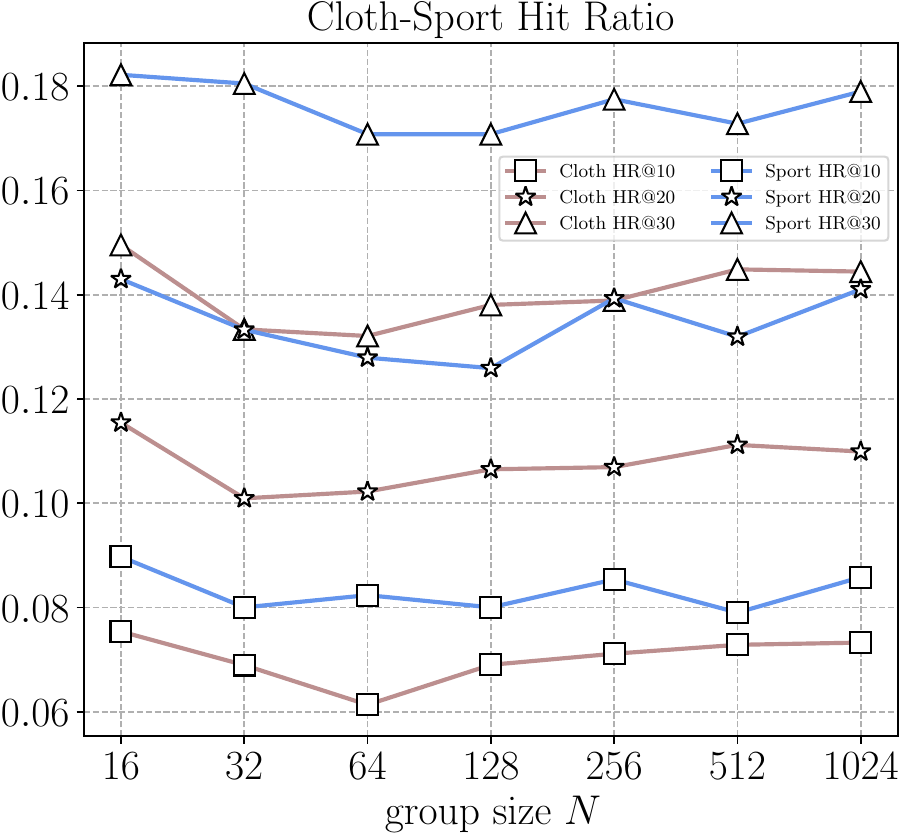}
        \label{fig:cs_groupsize}
    \end{subfigure}
    \begin{subfigure}[b]{.16\linewidth}
        \centering
        \includegraphics[width=\textwidth]{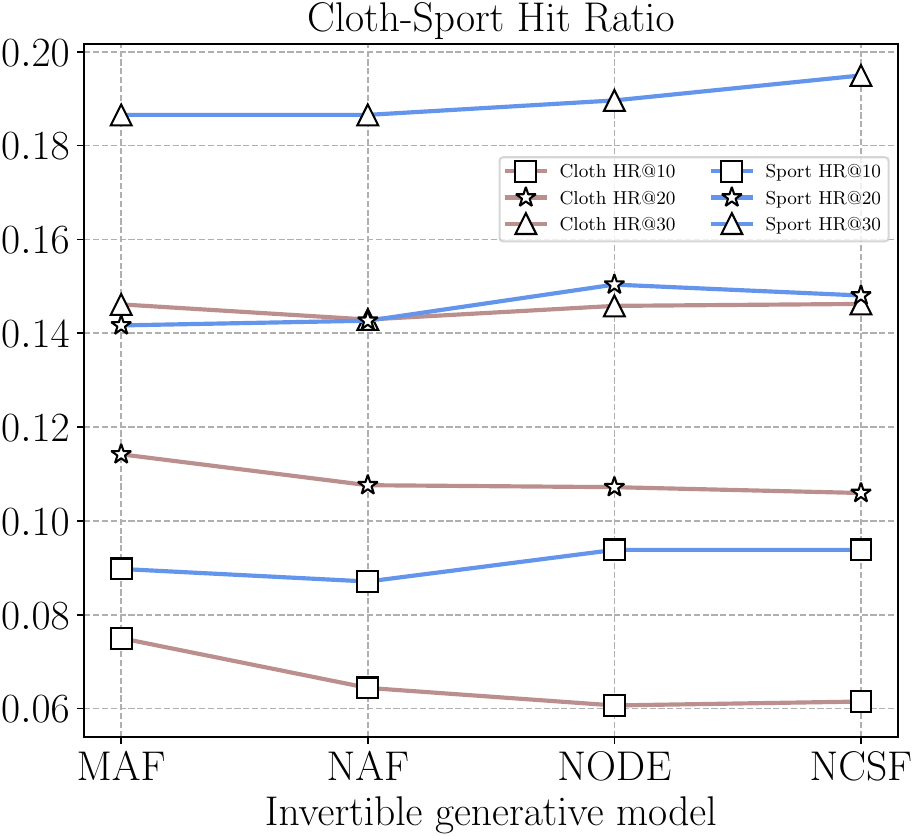}
        \label{fig:cs_flowtype}
    \end{subfigure}
    \begin{subfigure}[b]{.16\linewidth}
        \centering
        \includegraphics[width=\textwidth]{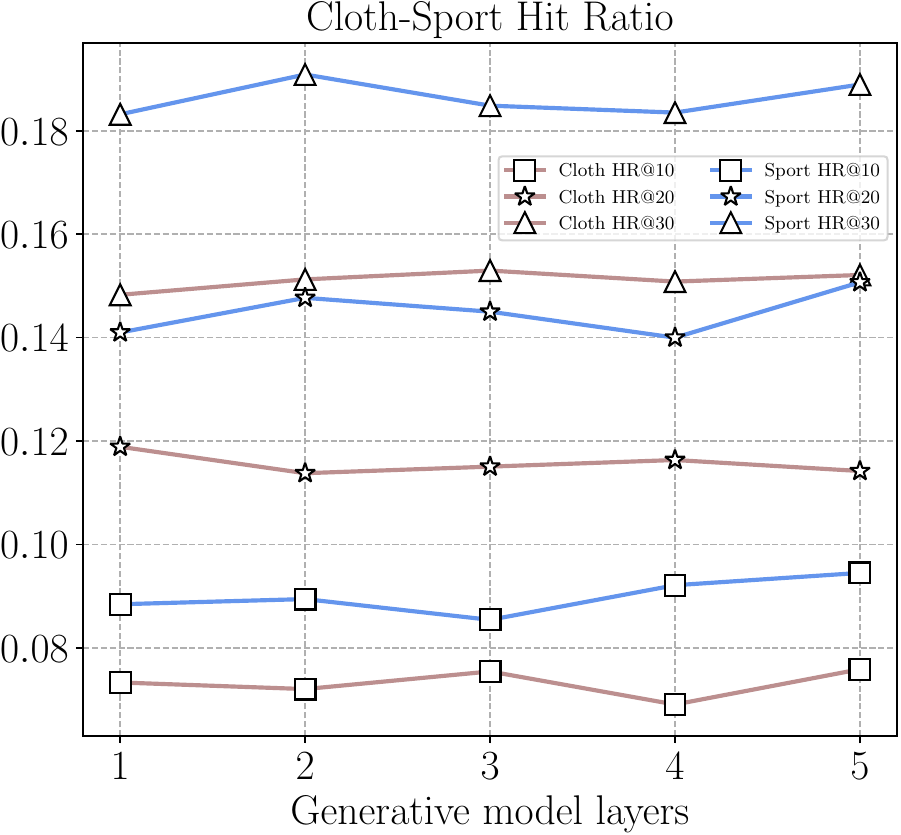}
        \label{fig:cs_flowlayer}
    \end{subfigure}
    \begin{subfigure}[b]{.16\linewidth}
        \centering
        \includegraphics[width=\textwidth]{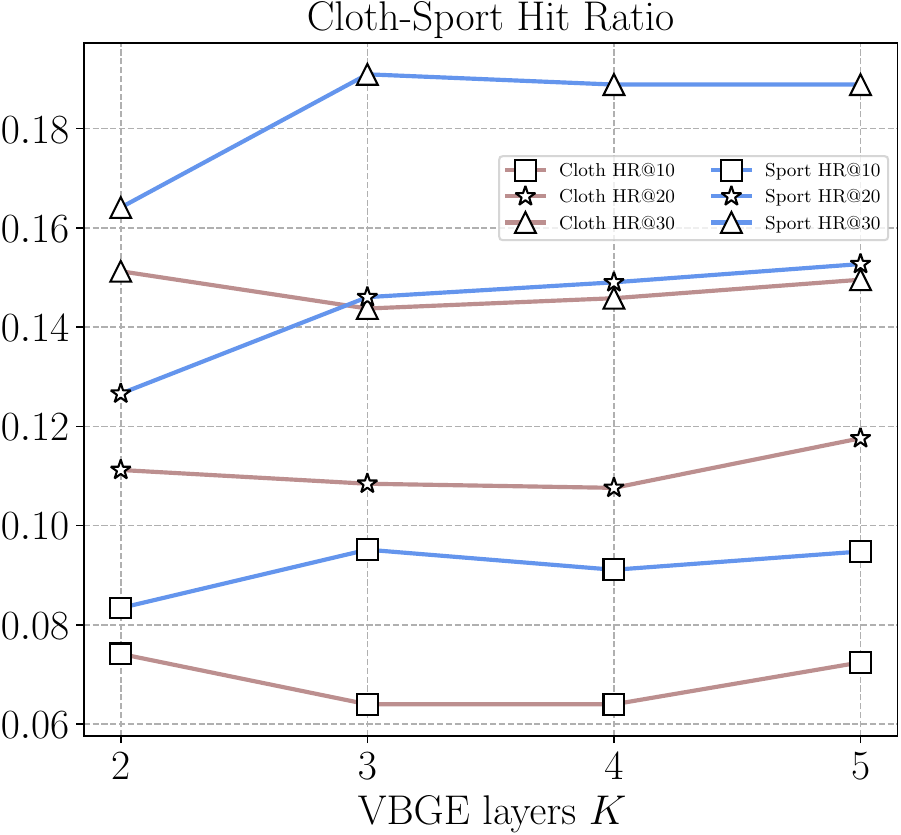}
        \label{fig:cs_align}
    \end{subfigure}
    \begin{subfigure}[b]{.16\linewidth}
        \centering
        \includegraphics[width=\textwidth]{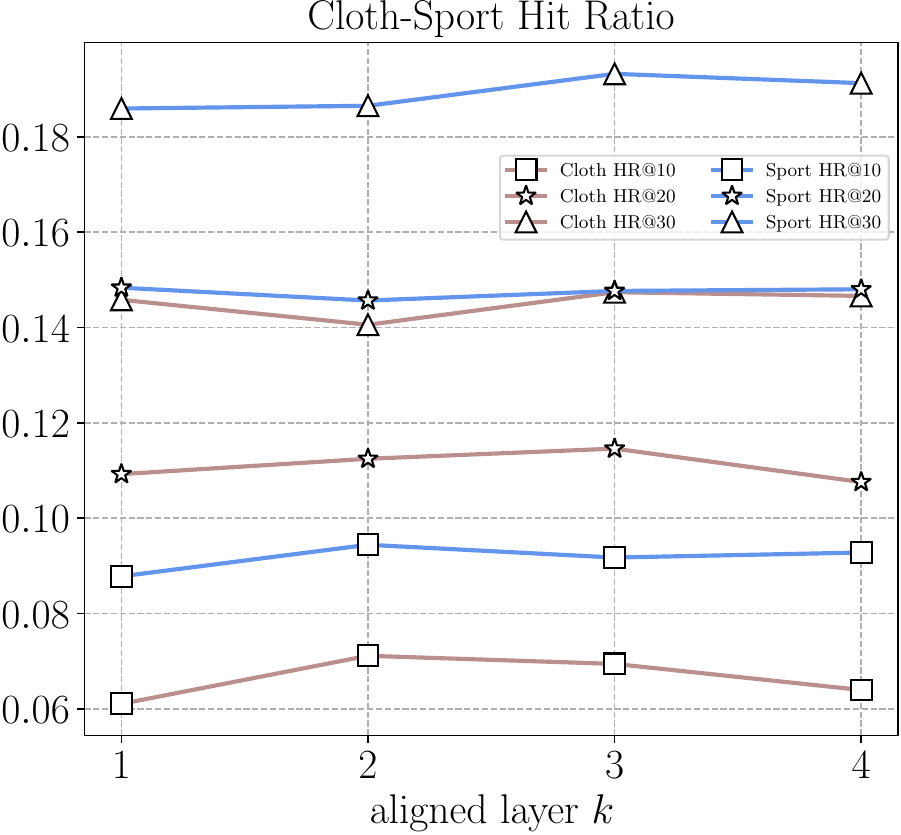}
        \label{fig:cs_gnnlayer}
    \end{subfigure}
    \begin{subfigure}[b]{.16\linewidth}
        \centering
        \includegraphics[width=\textwidth]{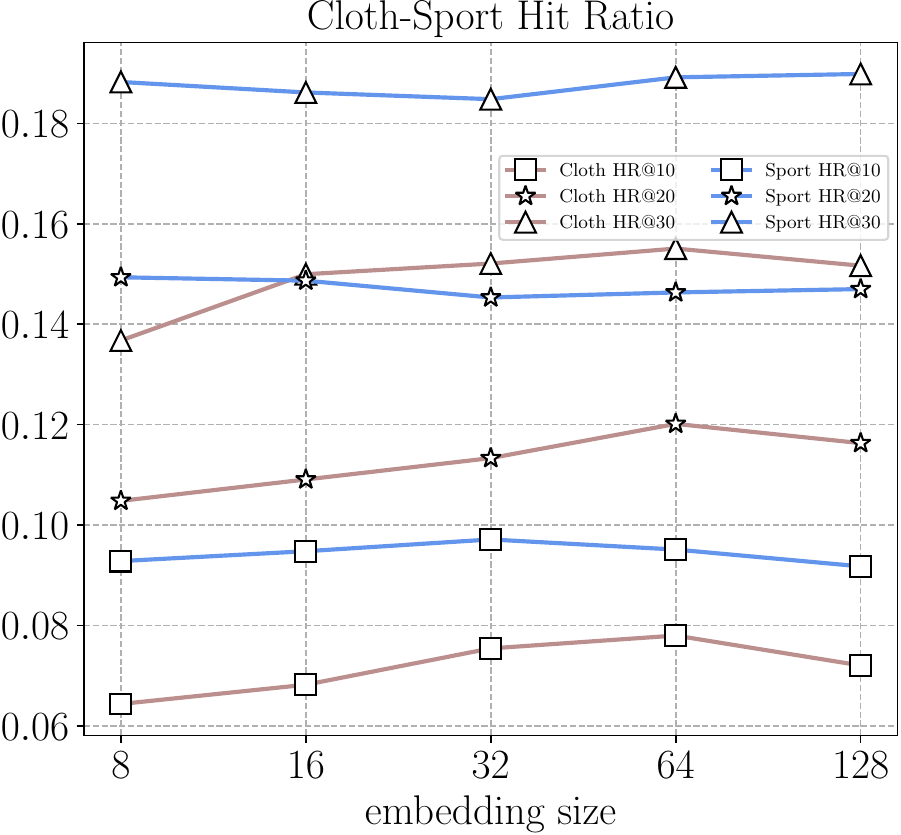}
        \label{fig:cs_embed}
    \end{subfigure}

    \begin{subfigure}[b]{.16\linewidth}
        \centering
        \includegraphics[width=\textwidth]{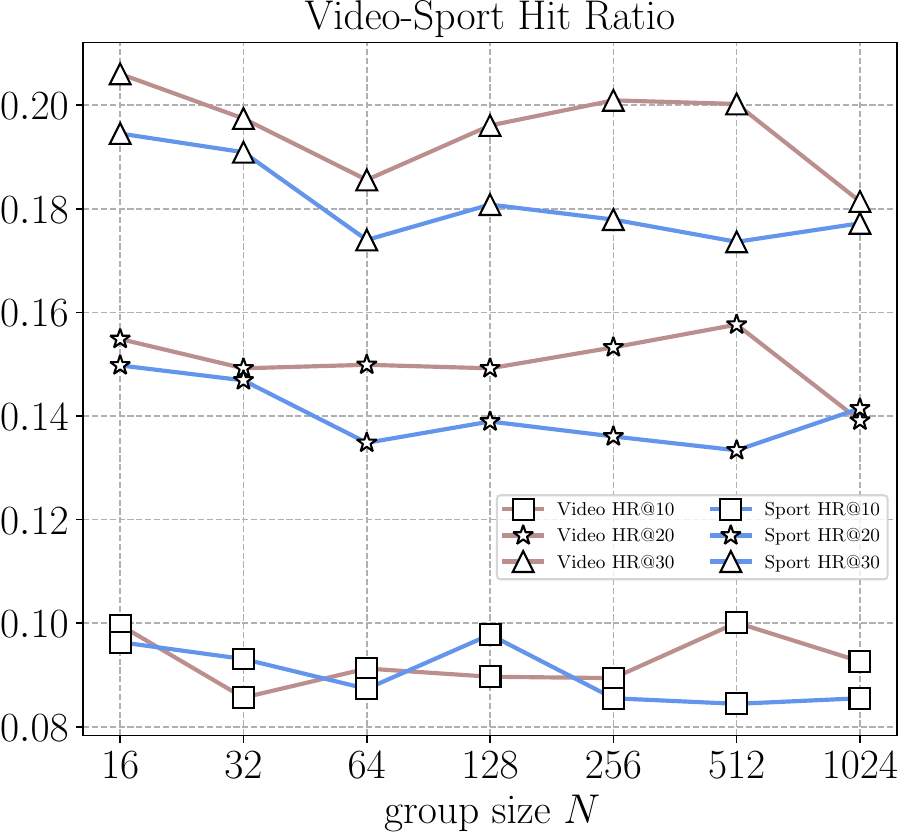}
        \label{fig:vs_groupsize}
    \end{subfigure}
    \begin{subfigure}[b]{.16\linewidth}
        \centering
        \includegraphics[width=\textwidth]{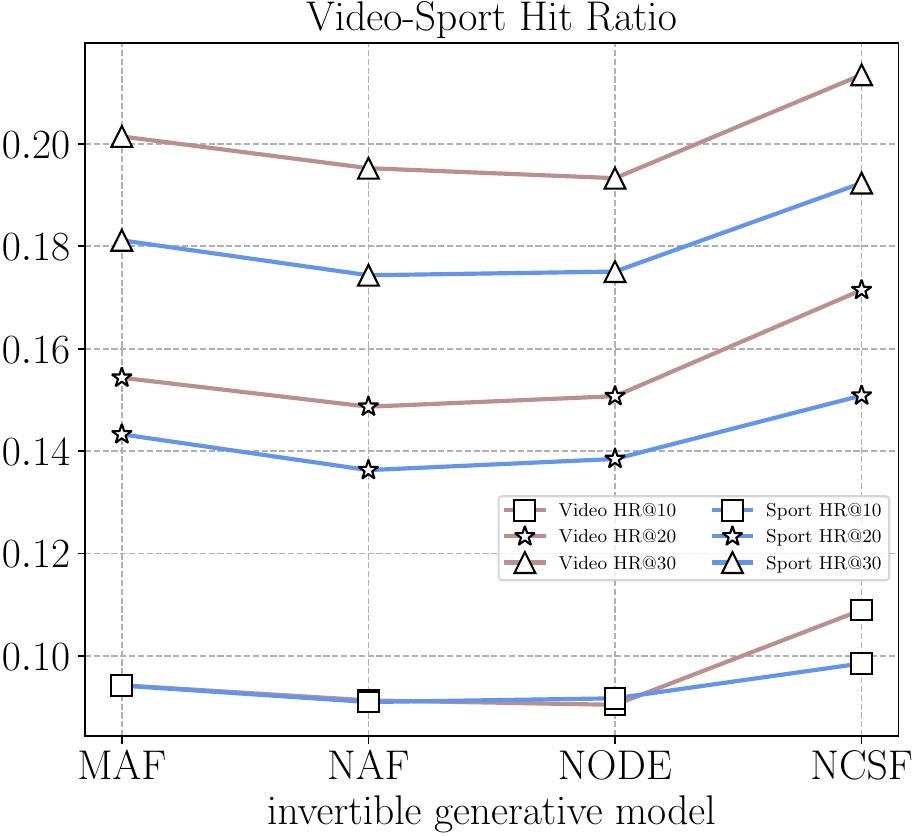}
        \label{fig:vs_flowtype}
    \end{subfigure}
    \begin{subfigure}[b]{.16\linewidth}
        \centering
        \includegraphics[width=\textwidth]{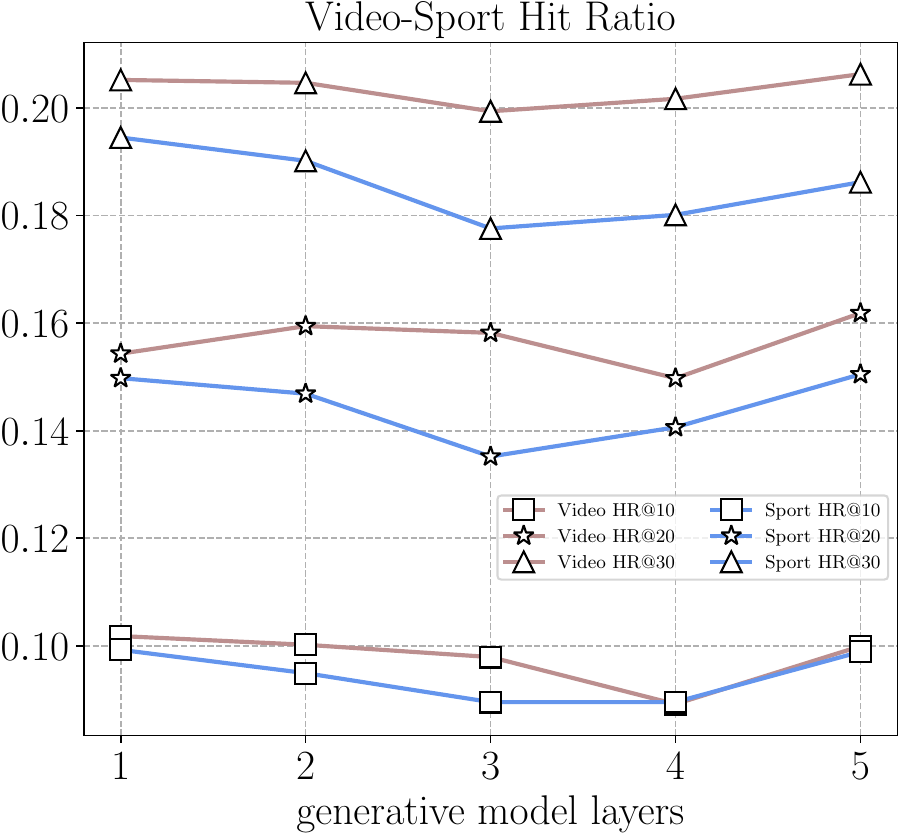}
        \label{fig:vs_flowlayer}
    \end{subfigure}
    \begin{subfigure}[b]{.16\linewidth}
        \centering
        \includegraphics[width=\textwidth]{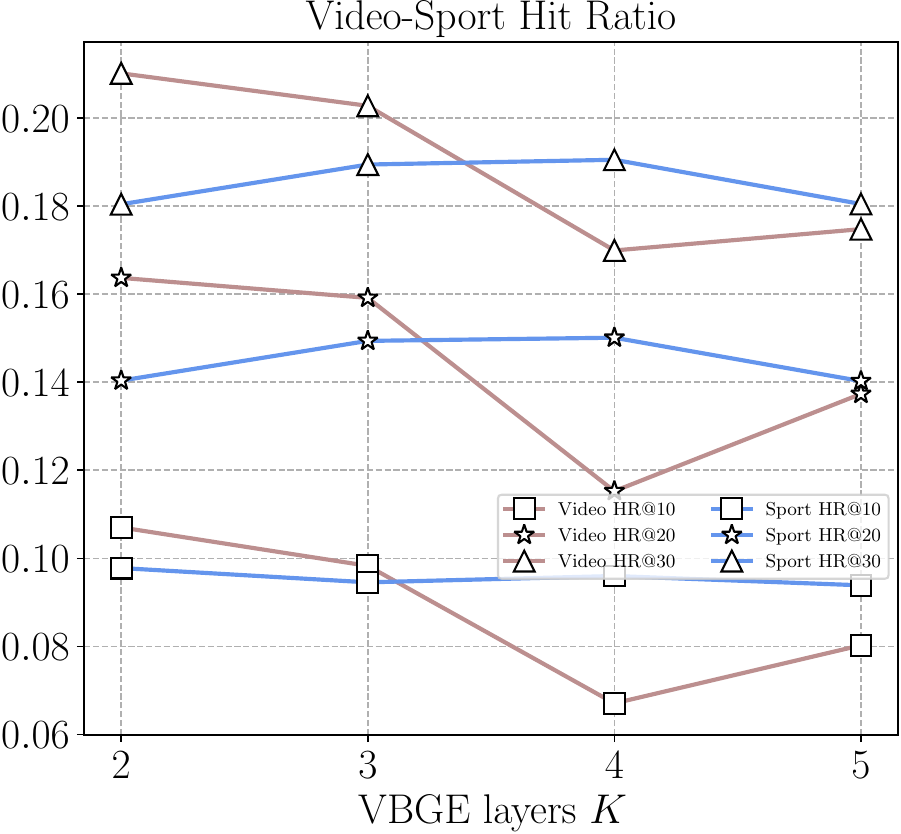}
        \label{fig:vs_align}
    \end{subfigure}
    \begin{subfigure}[b]{.16\linewidth}
        \centering
        \includegraphics[width=\textwidth]{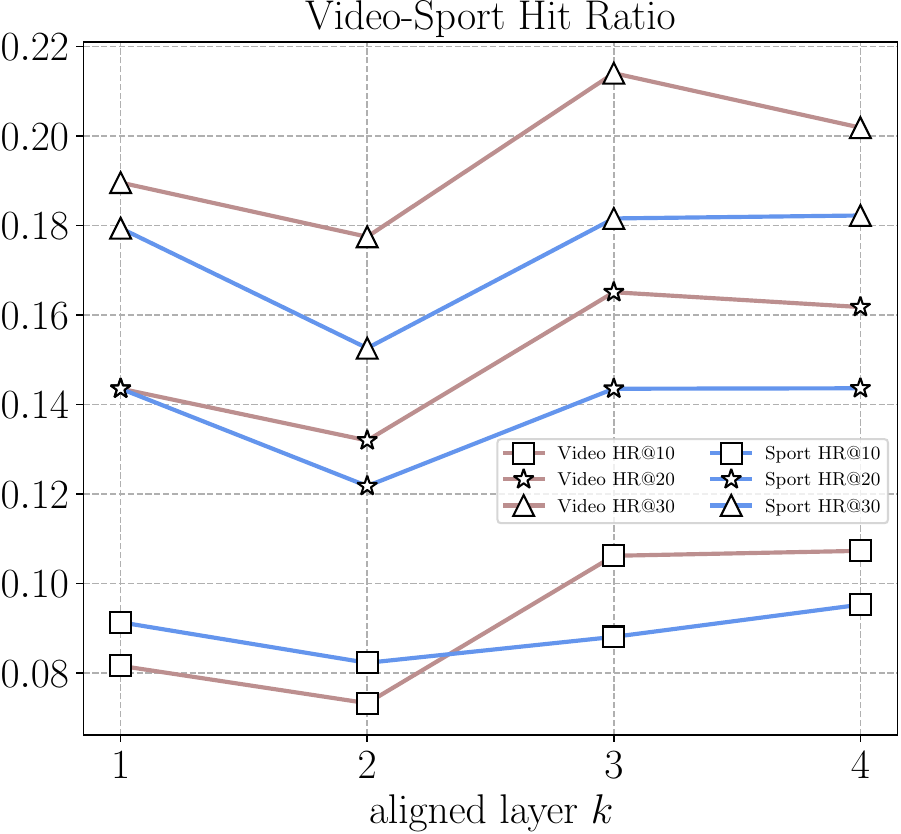}
        \label{fig:vs_gnnlayer}
    \end{subfigure}
    \begin{subfigure}[b]{.16\linewidth}
        \centering
        \includegraphics[width=\textwidth]{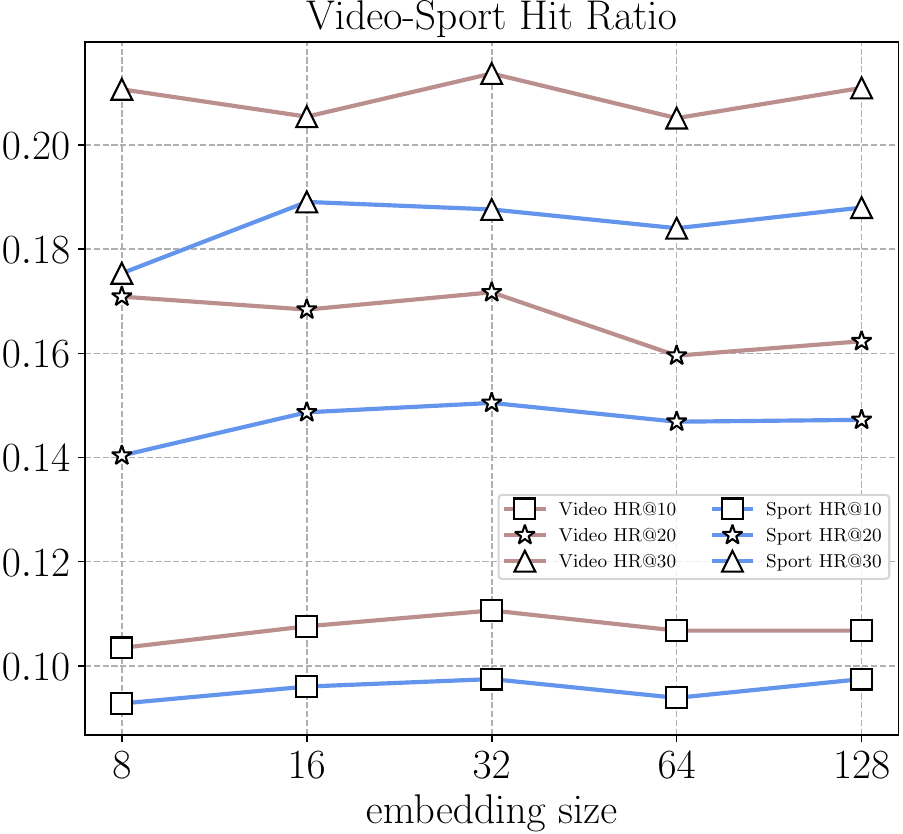}
        \label{fig:vs_embed}
    \end{subfigure}
    \vspace{-3mm}
    \caption{Parameter evaluation in 6 CDR tasks.}
    \label{fig: parameters}
    \vspace{-3mm}
\end{figure*}

We search the best value of the following parameters: user group size $N$, the invertible transformation and its chain length $L$, number of VBGE layers $K$, the depth of shallow layers $k$, and embedding size throughout the encoder neural network. The results in 6 CDR tasks are shown in Fig.~\ref{fig: parameters}.

\textbf{A small group size allows the best performance of HJID, with the runner-up results when the group size is large.}
We first evaluate the impact of \textbf{group size $N$}.
All metrics achieve the best performance when group size $N$=16.
A smaller group size allows our model to gradually approach the optimal solution through multiple iterations. 
We also observed a similar performance when the group size is sufficiently large. 
This could be attributed to the substantial number of random samples, enabling our model to fit the target joint distribution effectively.

HJID shows stable results when NCSF is used to generate variant parts in most tasks.
We then compare four \textbf{Flow-based models~(MAF, NAF, NODE, NCSF)} as the design choice of invertible transformations.
In the Game-Sport and Video-Sport tasks, HJID attains the best performance with NCSF, whereas NAF delivers the highest performance in the Video-Cloth and Game-Cloth tasks.
For a thorough comparison, we additionally conduct experiments with \textbf{varying layers} for both NAF and NCSF.
\textbf{The experimental results of 2-layer and 3-layer NCSF exhibit substantial similarity.}
In the Video-Sport task, 3-layer NAF performs the best, while in the Game-Sport task, 5-layer NCSF excels.
This emphasizes the importance of tailoring the model architecture according to the task at hand.

We extend our investigation to determine the optimal number of \textbf{VBGE layers $K$} in \{2, 3, 4, 5\}.
The single-layer VBGE is not considered, as HJID requires a minimum of two layers for proper implementation.
Empirical results indicate that \textbf{a 3-layer VBGE configuration yields superior performance across multiple metrics}.
Subsequently, we fix $K$=5 and examine the \textbf{most effective shallow layers $k$} for shallow subspace alignment within \{1, 2, 3, 4\}.
The optimal value of $k$ varies among different CDR tasks. 
\textbf{Higher aligned layers possess superior results}, encouraging HJID to decouple user representations into shallow-layer and deep-layer representations at higher levels.

Finally, we assess the \textbf{embedding size $d$} with 3-layer VBGEs and NCSF.
Since we concatenate the outputs of each layer, the size of representations spans across \{24, 48, 128, 192, 384\}.
\textbf{Optimal embedding sizes vary across different scenarios.}
The metrics exhibit an initial rise followed by a decline, with the optimal performance achieved at $d$=64 for Game-Sport and $d$=32 for Video-Cloth.

\vspace{-3mm}
\section{Related Work}
As of today, CDR methods fall into one of three categories: \textbf{Alignment} method~\cite{singh2008relational, agarwal2011localized, elkahky2015multi, gao2013cross}, \textbf{Bridge} method~\cite{man2017cross, zhu2018deep, kang2019semi, zhu2019dtcdr, bi2020heterogeneous, li2020ddtcdr, zhu2021transfer, zhu2022personalized, cao2022contrastive}, and \textbf{Constraint} method~\cite{cao2022disencdr, cao2022cross, du2023distributional, cao2023towards}.

For the \textbf{alignment method}, previous works share representations or align user representations in the latent space.
For example, CMF\cite{singh2008relational} is the first to factorize matrices of different domains simultaneously and directly share the embeddings of shared users across domains.
LFM\cite{agarwal2011localized} assumes that representations are generated from a user-specific global
latent factor.
CLFM\cite{gao2013cross} represents the cluster-level rating patterns of domains into a latent space. 
Also, MV-DNN\cite{elkahky2015multi} constructs a single mapping for user features and leverages information from interactions to a unified latent space.

As for \textbf{bridge methods}, their goal is to construct transformation functions across domains with bespoke neural architectures.
For example, EMCDR\cite{man2017cross} first proposes the embedding-and-mapping approach and applies the Multi-Layer Perception~(MLP) as the only transformation bridge to transfer information from the source to the target domain.
following, DCDCSR\cite{zhu2018deep}, SSCDR\cite{kang2019semi}, DTCDR\cite{zhu2019dtcdr}, HCDIR\cite{bi2020heterogeneous}, DDTCDR\cite{li2020ddtcdr}, TMCDR\cite{zhu2021transfer}, and PTUPCDR\cite{zhu2022personalized} obey the same procedure to capture cross-domain correlations between domains.
For example,
DCDCSR introduces DNN to generate latent factors across domains with the consideration of rating sparsity degrees.
HCDIR uses MLP to model the correlations between user representations in the latent space.
Moreover, 
TMCDR utilizes a meta-network to achieve good generalization in few-shot CDR, while PTUPCDR employs the meta-based network to generate personalized transformation bridges.
C$^2$DSR maximizes the mutual information between single-domain and cross-domain representations for unbiased information within and across domains.

However, these methods assumed a complete sharing of user representations across domains, ignoring the potential for negative transfer caused by non-transferable features.
To address this, recent researchers introduced various \textbf{constraints} to enhance the focus on sharable parts within user representations.
Specifically, DisenCDR\cite{cao2022disencdr} designs regularizers for overlapped and non-overlapped users to distinguish domain-shared and domain-specific information.
Similarly, CDRIB\cite{cao2022cross} incorporates variational information bottleneck into optimization to encourage the model to prioritize domain-shared information.
DPMCDR\cite{du2023distributional} summarizes user preferences as a domain-level preference distribution and aligns the invariant preferences of two domains using JS divergence.
Moreover, UniCDR\cite{cao2023towards} employs contrastive learning with domain masking mechanisms to constrain encoding domain invariant information.

Unlike previous methods that focus overwhelmingly on domain-shared information, our approach challenges their efficacy and disentangles domain-shared and domain-specific representations from the intertwined user representation and models the correlation between domain-specific factors and domain-shared factors in different domains in different domains based on data generation graphs, thus achieving CDR grounded in reasonable causal effect.
More importantly, our method ensures predictive robustness by establishing {\it identifiability} of the joint distribution across domains.

\vspace{-3mm}
\section{Conclusion and Future Works}
We have presented a causal subspace disentanglement method that resolves issues in identifying cross-domain joint distribution, while maintaining cross-domain consistency and domain-specific properties.
Particularly, we divided user representations into shallow general features and deep domain-oriented features, according to the Hierarchical Feature Representation principle. 
As a way to align domain-irrelevant general features, we employed Maximum Mean Discrepancy to domain-shared representations in shallow subspace.
We then ensured the {\it identifiability} of cross-domain joint distributions by building a causal data generation graph with disentangled deep-layer latent variables.
Empirical assessments in the real world show HJID's superior performance over existing methods, establishing a new state-of-the-art. Future work will focus on uncovering cross-domain causal relationships and model generalization to multi-domain contexts.

\clearpage
\bibliographystyle{ACM-Reference-Format}
\bibliography{Reference.bib}

\end{document}